\documentclass[twocolumn,astrosymb,usenatbib]{aastex7}
\usepackage{xcolor}
\usepackage{amsmath}	
\usepackage{amssymb}	
\usepackage{physics}   
\usepackage{verbatim}

\usepackage{float}

\defcitealias{Contigiani2019}{C19}
\defcitealias{Gallo2022}{G22}

\begin{document}

\title[Non-axisymmetry, time dependence, and HVS trajectories]{The impact of the Galactic bar and the Large Magellanic Cloud on hypervelocity star trajectories}

\author[0009-0006-1587-719X]{Isabella Armstrong}
\affiliation{Department of Physics and Astronomy, McMaster University, 1280 Main Street West, Hamilton, ON, L8S 4M1, Canada}
\email[show]{armstri@mcmaster.ca}

\author{Fraser A. Evans}
\affiliation{David A. Dunlap Department of Astronomy and Astrophysics, University of Toronto, 50 St. George Street, Toronto, ON, M5S 3H4, Canada}
\affiliation{Dunlap Institute for Astronomy \& Astrophysics, University of Toronto, 50 St. George Street, Toronto, ON, M5S 3H4, Canada}
\email{fraser.evans@utoronto.ca}

\author{Jo Bovy}
\affiliation{David A. Dunlap Department of Astronomy and Astrophysics, University of Toronto, 50 St. George Street, Toronto, ON, M5S 3H4, Canada}
\affiliation{Dunlap Institute for Astronomy \& Astrophysics, University of Toronto, 50 St. George Street, Toronto, ON, M5S 3H4, Canada}
\email{jo.bovy@utoronto.ca}

\begin{abstract}

Hypervelocity stars (HVSs) ejected from the Galactic Center (GC) at speeds faster than the Galactic escape velocity are useful tools to provide insight into the Milky Way's dark matter halo. However, most characterizations of HVS orbits assume static models of the Milky Way's gravitational potential. In this work, we assess the influence of the Galactic bar and the Large Magellanic Cloud (LMC) on HVS trajectories, comparing them with those from an axisymmetric potential. We simulate 28,000 HVSs ejected over the last 100 Myr and find that ignoring the bar and LMC can cause their apparent ejection location to drift by up to 100 pc. Applying two standard HVS potential fitting methods to our sample shows that they are unable to perform as designed when non-axisymmetric effects are neglected. We calculate the angle between HVS Galactocentric position and velocity and find the LMC and bar can induce a deflection angle of up to several degrees. Using mock Gaia Data Release 4 observations, however, we show that this deflection is too small in magnitude to be measured in the near future without significantly improved observational uncertainties, particularly in heliocentric distance. Our results emphasize the need to account for the bar and LMC in modeling the Galactic potential using HVSs as a tracer.

\end{abstract}

\keywords{Hypervelocity stars (776) -- Milky Way dark matter halo (1049) -- Galactic bar (2365) -- Large Magellanic Cloud (903) -- Galactic Center (565) -- supermassive black holes (1663) }

\section{Introduction}\label{sec:intro}

The discovery of the first hypervelocity star (HVS) candidate with a heliocentric radial velocity of 850 $\mathrm{km \ s^{-1}}$ by \cite{Brown2005} provided the first observational evidence for the population of fast-moving stars predicted by \cite{Hills1988}. In subsequent years, further observational efforts uncovered more stars traveling at unusually high velocities \citep{Hirsch2005, Edelmann2005, Brown2006, Brown2007, Brown2014, Huang2017, Irrgang2019}, exceeding the Milky Way's escape velocity, which is about 500 km/s near the Sun \citep{Koppelman2021}. One recent example is the star S5-HVS1, traveling at 1700 km/s in the rest frame of the Galaxy \citep{Koposov2020} \par 


There are several proposed ejection mechanisms for these stars, most notably the \cite{Hills1988} mechanism in which a stellar binary is tidally disrupted by Sgr A*, the $4 \times 10^{6} \ \mathrm{M_\odot}$ supermassive black hole \citep{Ghez2008, Akiyama2022, GRAVITY2018} located in the Galactic center (GC). One member of the former binary is captured in orbit around the MBH and its former companion is ejected as an HVS. Here, we focus specifically on HVSs ejected from the GC in this way at velocities unbound from the Galaxy. However, other proposed ejection mechanisms \textit{not} located at the GC include the ejection of a companion following a supernova event in a stellar binary, either core collapse \citep{Blaauw1961, Portegies2000, Evans2020} or thermonuclear \citep{Geier2013, Geier2015, Shen2018, Tanikawa2019, Neunteufel2021}, or as the result of dynamical interactions of massive stars in dense clusters \citep{Poveda1967, Perets2012, Oh2016, Capuzzo2015}. 



Hypervelocity stars have been proposed as a promising tracer population to study the gravitational potential of the Milky Way \citep{Gnedin2005, Yu2007, Contigiani2019, Gallo2022}. The use of test particle kinematics to determine the potential consistent with those motions is called \textit{dynamical modelling} has a long history in classical mechanics and Milky Way astronomy in particular, using a variety of tracers including globular clusters \citep{Eadie2019, Posti2019}, streams \citep{Reino2021, Koposov2023}, Milky Way satellite galaxies \citep{Callingham2019, Slizewski2022, Li2025}, halo stars \citep{Deason2021} and stars near the Galactic escape velocity \citep{Prudil2022, Roche2024}. The basic issue at hand is that accelerations in the Galaxy are small in magnitude and quite difficult to measure over human timescales. Even assuming positions and velocities of tracer populations can be measured to arbitrary precision, they cannot constrain the Galactic potential alone, as the potential determines the acceleration of test particles but not the positions and velocities directly. Inferring the potential from tracer kinematics therefore requires additional assumptions and introduces significant caveats and degeneracies into the derived constraints. The required assumptions and statistical technique(s) most appropriate to each tracer are explored in depth in textbooks and recent reviews \citep{BinneyTremaine2008, Bland-Hawthorn2016, Wang2020, deSalas2021, Gardner2021, Hunt2025}. 

An important aspect of HVSs that can be leveraged to constrain the potential is that once identified, they are known to have been located at the Galactic Centre at some point in the past. This means that along with their current position and velocity, the spatial component of a second phase space measurement is available for `free', and this second phase-space position is likely well-separated from the first in time. 



The Milky Way potential is commonly modeled using a spherical bulge component, one or more axisymmetric disc components, and a spheroidal dark matter (DM) halo \citep[e.g.][]{Irrgang2013, Bovy2015, McMillan2017, Cautun2020}. 
Such potentials are overall axially symmetric in the plane of the disc -- the trajectory of an HVS traveling through a potential model comprised of these components should not be deflected at all in the azimuthal direction and should only be deflected in the polar direction due to the presence Galactic disc and the DM halo (provided it is oblate or prolate). In this simplified picture, methods to constrain the Galactic DM halo using HVSs have been proposed \citep[e.g.][]{Contigiani2019, Gallo2022}. However, the Milky Way potential contains known asymmetric components including, among others, the Large Magellanic Cloud (LMC) and the Galactic bar. The Galactic bar introduces asymmetry only in the azimuthal direction, while the infalling LMC introduces asymmetry in both the polar and azimuthal directions. Additionally, the LMC also introduces non-equilibrium by inducing a displacement of the Milky Way barycenter over time, hereafter referred to as the \textit{reflex motion} or \textit{non-inertial frame} force \citep[NIF;][]{Penarrubia2016, GaravitoCamargo2019, Petersen_2020}. 

Previous works have shown that the LMC and its associated reflex motion appreciably influence HVS trajectories \citep{Kenyon2018, Boubert2020}. This work aims to expand this discussion by additionally studying the impact of the Galactic bar. 
Using numerical simulations of HVS populations, we examine how the Galactic bar, LMC, and reflex motion cause HVSs to deviate from trajectories expected from a typically assumed axisymmetric potential. We apply the proposed potential fitting schemes of \citet{Contigiani2019} and \citet{Gallo2022} to determine whether the effects of the LMC and Galactic bar are significant when using HVS trajectories to infer the Galactic potential.
Finally, we obtain mock photometric, astrometric and radial velocity measurements from the European Space Agency's Gaia mission \citep{Gaia2016} to determine what fraction of HVSs are expected to be observable in the near future and if we can measure the deflection caused by non-axisymmetry and time dependence in the Milky Way's potential.

This paper is structured as follows. We discuss methodology and the parameters of our simulations in Section \ref{sec:methods}, outlining how our HVS populations are generated, the construction of the different Galactic potentials, the cuts imposed upon the sample, the potential fitting methods, and how mock Gaia observations were performed. In Section \ref{sec:results} we discuss our results. We talk about how the inclusion of the bar, LMC and reflex motion impact the trajectories of HVSs (Sec. \ref{sec:results-traj}), their application to potential fitting (Sec. \ref{sec:results-potential}), how these components deflect HVSs compared to an axisymmetric Milky Way potential  (Sec. \ref{sec:results-deflect}), and where this effect is observable (Sec. \ref{sec:results-obsv}). 
We discuss the significance of these results in Section \ref{sec:discussion} and conclude by summarizing key results in Section \ref{sec:conclusions}. 

\section{Methodology} \label{sec:methods}
We look at the trajectories of HVSs by propagating them both forward and backwards in time through our potentials of interest. In this section, we outline how we generate our synthetic HVS populations, define potential components, and implement the potential constraining methods of \cite{Contigiani2019} and \cite{Gallo2022}. 


There are five different variations of the Milky Way potential we are interested in, each of which we give a shorthand label for reference: 

\begin{enumerate}
    \item A fiducial Milky Way composed only of axisymmetric components (MW-only).
    \item A Milky Way potential including the additional non-axisymmetric bar component  (MW+Bar).
    \item A non-axisymmetric Milky Way potential including the LMC, neglecting the reflex motion of the GC (MW+LMC). 
    \item A non-axisymmetric Milky Way potential set in the non-inertial reference frame induced by the reflex motion of the GC, but \textit{not} including the actual gravitational influence of the LMC itself (MW+NIF).
    \item A non-axisymmetric Milky Way potential including the LMC, reflex motion of the GC, and Galactic bar (MW+All).
\end{enumerate}


\subsection{Initial conditions}\label{sec:methods:initial}

We use the Python package \texttt{speedystar}\footnote{\url{https://github.com/fraserevans/speedystar}} to generate initial conditions for a population of HVSs ejected via the \cite{Hills1988} mechanism  \citep[for details see][]{Evans2022b}. The size of the ejection catalog is determined by the ejection rate of HVSs from Sgr A*, with a mass/velocity distribution using a Monte Carlo approach \citep[see also][]{Marchetti2018, Evans2021, Evans2022a}. We assume an HVS ejection rate of $10^{-4} \; \mathrm{yr^{-1}}$, yielding a catalog sufficiently large to investigate the bar and LMC impacts -- the true ejection rate is likely at least an order of magnitude smaller \citep{Evans2022a, Evans2022b, Verberne2024}. HVS progenitor binary population adheres to the \texttt{speedystar} defaults.
Each star is initialized 3 pc from Sgr A*, i.e. the edge of its sphere of influence \citep{Sari2010, Kobayashi2012, Rossi2014}. Position is drawn isotropically, with velocity pointed radially away from the GC. 
The star is assigned a random flight time $t_{\rm flight}$, which is the time it will travel for after being ejected
Maximum flight time is limited to 100 Myr since HVSs ejected longer ago are overwhelmingly likely be too far away to be detectable in current or near-future Galactic surveys. HVSs are evolved using the single stellar evolution \citep[SSE;][]{Hurley2000} relations as implemented within \texttt{AMUSE} \citep{Portegies2013, Pelupessy2013, Portegies2018}.We assume each HVS has spent a random fraction of its lifetime (the time between the zero age main sequence and the beginning of the asymptotic giant branch) before being ejected. Stars which are too old at their ejection time to survive for their flight time as a non-remnant until the present day are removed from the sample. From these stellar evolution prescriptions we get the size, luminosity and effective temperature of each HVS, allowing for mock observations of our sample.



\subsection{The Milky Way Gravitational Potential} \label{sec:methods:potential}
Using this ejection catalog, each individual star is propagated forward in time using \texttt{galpy}\footnote{\url{https://github.com/jobovy/galpy}~.} \citep{Bovy2015}. This allows HVSs with the same initial conditions (initial position, flight time, etc) to be propagated through different potentials. In this work, the ejection catalog of stars is propagated through each of the five cases of interest (MW, MW+Bar, MW+LMC, MW+NIF, MW+All).
This allows us to directly compare the impact of each individual potential across the entire HVS catalog. 

We begin by defining the components of our fiducial MW-only potential. We use the \texttt{MWPotential2014} potential of \citet{Bovy2015}, modelled on various kinematic data \citep[see also][]{Bovy2013}, which consists of a bulge described as a power law with an exponential cutoff, a disk modelled as a \citet{Miyamoto1975} potential, and a dark matter halo modelled as an NFW profile \citep{Navarro1997}. For consistency, we add to this potential an extra contribution from Sgr A*, described simply as a point mass of $4\times 10^6 \; \mathrm{M_\odot}$ \citep{Ghez2008}. The qualitative results of this study are not sensitive to reasonable variations of this fiducial potential -- our conclusions remain the same under other commonly used axisymmetric Milky Way potentials \citep[e.g. those described in][]{Irrgang2013, McMillan2017, Cautun2020} .

The above components describe our fiducial axisymmetric Milky Way potential (MW-only). In potentials that include the Galactic bar (MW+Bar and MW+All), the contribution of the Galactic bar is modeled using a \citet{Dehnen2000} bar potential which has been generalized to three dimensions following \cite{Monari2016}:
\begin{align}\label{eqn:Dehnen}
    \Phi(R,z,\phi) = &A_f\cos(2[\phi-\Omega_bt])\left(\frac{R}{r}\right)^2  \\ & \ \times\begin{cases}
        -(R_b/r)^3, &\,\text{for }\, r\geq R_b \\
        (r/R_b)^3-2, &\,\text{for }\, r\leq R_b \; ,
    \end{cases}
\end{align}
where $r^2 = R^2 + z^2$ is the spherical radius. We assume the major axis of the bar is offset from the line connecting the Sun and the GC by 25$^\circ$ in the present day, a bar speed of $\Omega_b = 40 \; \mathrm{km \ s^{-1} \ kpc^{-1}}$ \citep{Sormani2015, Bovy19,Leung23}, a bar radius of $R_b =5$ kpc and a dimensionless bar strength of $A_f = 660$ \citep{Wegg2015, Portail2015, Portail2017}. We assume the bar has been rotating with a fixed speed, size and strength over the range of HVS flight times we explore. It is important to note that this formalism of the bar is not an additional mass component superimposed upon the others, rather it redistributes existing mass to be denser in regions where the bar exists. 
\par 

For potentials which include the gravitational presence of the moving LMC (MW+LMC and MW+All), the LMC is modeled using a \citet{Hernquist1990} potential with a mass of $M_{\rm LMC} = 1.5 \times 10^{11}$M$_{\odot}$ and a radius of $r_{\rm LMC} = 17.14$ kpc. \citep{Erkal2019LMCmass, Erkal2020LMCmass}. The LMC is initialized at time $t=0$ at its present day center of mass position of RA = $+78.76^\circ$, DEC = $-69.19^\circ$ and distance of 49.59 kpc \citep{Zivick2019, Pietrzynski2013}, proper motions of $\mu_{\rm RA}^* = +1.91 \; \mathrm{mas \ yr^{-1}}$, $\mu_{\rm DEC} = +0.229 \; \mathrm{mas \ yr^{-1]}}$ and heliocentric line-of-sight velocity +262.2 km s$^{-1}$ \citep{vanderMarel2002, Kallivayalil2013}. We then integrate the LMC orbit backwards in time over the last 100 Myr through the MW-only potential including the effect of dynamical friction \citep{Chandrasekhar1943}. We use \texttt{galpy}'s \texttt{MovingObjectPotential} option to calculate its contribution to the Galactic potential with time.  

For potentials which include the impact of the reflex motion (MW+NIF and MW+All), a non-inertial reference frame is created by calculating the forces the LMC exerts on the Galactic barycenter. Using the computed LMC orbit, the location of the GC is explicitly set, and the force of acceleration on the GC due to the moving LMC is calculated in the $x$, $y$, and $z$ directions. These forces are then interpolated over the entire time domain, spanning from 100 Myr ago to the present day. The non-inertial force acting on the frame can then be treated as a component of the Galactic potential. This process is detailed in the \texttt{galpy} documentation\footnote{\url{https://docs.galpy.org/en/v1.10.0/orbit.html\#orbit-example-barycentric-acceleration-lmc}}.

To compute the trajectory of a star with flight time $t_{\rm flight}$ Myr, each of the time-varying components of the potential are `wound back' in time to $t_{\rm flight}$ Myr ago. The star is then integrated from $t=-t_{\rm flight}$ until the present day at $t=0$ in the time-varying Milky Way potential in \texttt{galpy} using a Dormand-Prince integrator \citep{Dormand1980} with a time step of 0.05 Myr. This results in a simulated population of HVSs located where we would expect to see them today had they been ejected $t_{\rm flight}$ years ago.  

Most results in this work are shown in Galactocentric Cartesian coordinates, but when relevant we assume the Galactic circular velocity at the Solar circle is 233.4 $\mathrm{km \ s^{-1}}$ \citep{Drimmel2018} and that the Sun is 8.122 kpc from the GC \citep{GRAVITY2018}, 20.8 pc above the Galactic midplane \citep{Bennett2019} and has a velocity relative to the Local Standard of Rest of $[U_\odot,V_\odot, W_\odot]=[11.1, 12.24, 7.25] \; \mathrm{km \ s^{-1}}$ \citep{Schonrich2010}.  

\subsection{Selecting the HVS sample}\label{sec:methods-select-sample}


From our five catalogs of interest (MW-only, MW+Bar, MW+LMC, MW+NIF, MW+All), we select stars with Galactocentric velocity greater than the Galactic escape velocity at their current position. This is given by
\begin{align}
    v_{esc} = \sqrt{2(\Phi(\infty) - \Phi(R,z)} \; ,
\end{align}
where $\Phi(\infty)$ is evaluated at $R = 1$ Gpc, $z = 0$ kpc. This ensures that only HVSs that are gravitationally unbound from the Milky Way are selected. After removing stars which do not survive until the present day and selecting only unbound stars, we are left with five catalogs of $\sim$23,000 HVSs each.


With these catalogs in hand, it is useful to know which of the stars would be detectable in the near future and the precision to which their kinematics could be measured. To calculate the mock magnitudes of these stars in the Gaia photometric bands and obtain mock astrometric errors, we follow the approach used in \cite{Evans2022b}. Using the effective temperatures, luminosities and surface gravity of each HVS along with the visual extinction along its line of sight (using the \texttt{combined15} Galactic dust map of \citealt{Bovy2016}), we estimate each mock HVS's apparent magnitude in the Gaia and Johnsons-Cousins bands by interpolating the \texttt{MIST} model grids \citep{Choi2016, Dotter2016}. We additionally obtain mock magnitudes in the Gaia $G_{\rm RVS}$ band for each star using their Gaia $G$ and Johnson-Cousins $V$ and $I_c$ magnitudes and the polynomical fits of \citet{Jordi2010}. Given that no high-confidence HVS candidates have yet been unearthed in the most recent Gaia Data Release 3 \citep{Marchetti2022}, we instead project forward in time to the forthcoming Data Release 4 (DR4). To appear in the radial velocity subset of DR4, an HVS must be sufficiently bright to appear in the catalog ($G_{\rm RVS}<16.2$ for stars cooler than 6900 K, $G_{\rm RVS}<14$ otherwise) To estimate astrometric errors, we use the DR3 astrometric spread function of \citet{Everall2021cogiv}, which yields the full five-dimensional astrometric covariance matrix for a source. To mimic DR4 uncertainties, we scale down the estimated DR3 astrometric errors according to the expected improvement from DR3 to DR4\footnote{\url{ https://www.cosmos.esa.int/web/gaia/science-performance}, \citet{Brown2019}}. We estimate radial velocity errors for these sources using the package \texttt{PyGaia}\footnote{\url{https://github.com/agabrown/PyGaia}} based on the stellar effective temperature and RVS-band magnitude.

\subsection{Constraining the Galactic potential}\label{sec:methods-likelihood}

We aim to explore how non-axisymmetry and time dependence in the potential impacts how HVSs can be used as a dynamical tracer for the Galactic potential, in particular the Galactic DM halo. If HVSs have traveled through a potential including a bar and/or LMC but their orbits are used to fit to a simpler potential model, can meaningful and accurate constraints still be made? To explore this, we turn to two different formalisms outlined in \citet{Contigiani2019} and \citet{Gallo2022}, hereafter \citetalias{Contigiani2019} and \citetalias{Gallo2022}, respectively, for constraining the Galactic DM halo using HVS observations. We summarize the approaches here and refer the interested reader to the papers for more detailed information. 

\subsubsection{The trajectory method}

Consider an HVS in the present day at time $t>0$ which is located somewhere in the seven-dimensional configuration space $\textbf{w}=(\textbf{x}, \textbf{v}, m)$, where $(\textbf{x}, \textbf{v})$ are the three-dimensional position and velocity vectors and $m$ is the HVS mass. The HVS was ejected from the GC with some initial configuration $\textbf{w}_i=(\textbf{x}_i, \textbf{v}_i, m)$ at a time $t_i < t$, assuming that the mass does not change with time. The distribution of possible initial configurations is described by $R(\textbf{w}_i)$, with units of configuration space density per unit time. In the case of the Hills mechanism, for example, this density is largest in regions of configuration space corresponding to positions close to the GC.
We will define a more explicit form for $R(\textbf{w}_i)$ shortly.

We can define the HVS distribution function $f(\textbf{w},t) \equiv \frac{dN(t)}{d^7\textbf{w}}$ by integrating over all possible ejection times and initial configurations:
\begin{equation}
\begin{split}
    f(\textbf{w},t) = \int d^7 \textbf{w}_i \int_0^t dt_i &R(\textbf{w}_i) \times g(t-t_i,m) \\
    &\times \delta(W(\textbf{w}_i,t,t_i)-\textbf{w}) \; \text{,}
\end{split}
\end{equation}
where $W(\textbf{w}_i, t, t_i)$ is the solution to the equations of motion which map the initial configurations $\textbf{w}_i$ at time $t_i$ to $\textbf{w}$ at time $t$. The survival function $g \leq 1.0$ takes into account the finite lifetime of stars, since not all stars ejected at $t_i$ will necessarily survive until $t$ as a non-remnant. For simplicity, however, we assume $g$=1.0 over all $(t-t_i,m)$ as it does not qualitatively impact our results. Performing the integral over the Dirac delta,
\begin{equation}
    f(\textbf{w},t) = \int_0^t dt_i R(W_i(\textbf{w},t,t_i)) \; \text{,}
\end{equation}
where $W_i(\textbf{w},t,t_i)$ is the solution to the delta function, i.e. the initial configuration at time $t_i$ which corresponds to final configuration $w$ at time $t$, which can be obtained by integrating the equations of motion backwards in time from $\textbf{w}$. Since the appropriate $W_i$ depends on the assumed Galactic potential, $f(\textbf{w},t)$ depends on the potential as well. For a potential defined by a set of parameters $\Theta$, we can make this dependence explicit by rewriting $f(\textbf{w},t)$ as
\begin{equation}
    f(\textbf{w},t|\Theta) = \int_0^t dt_i R(W_i(\textbf{w},t,t_i|\Theta)) \; \text{.}
\end{equation}

Given a sample of $N_{\rm HVS}$ stars, a likelihood can be assigned to any given potential:
\begin{equation} \label{eq:L}
    \mathcal{L}(\Theta) = \sum_{j=1}^{N_{\rm HVS}} \mathcal{L}_j(\Theta) = \sum_{j=1}^{N_{\rm HVS}} f(w_j,t|\Theta) \; \text{,}
\end{equation}
where $\textbf{w}_j$ is the coordinate in configuration space of the $j$'th HVS in the sample. The Galactic potential which best fits the set of HVS observations is that which maximizes $\mathcal{L}(\Theta)$ by way of maximizing $R(W_i(\textbf{w},t,t_i|\Theta))$ within the integral.

In plainer language, the above can  be summarized by simply saying that the Galactic potential with the maximum likelihood is the potential under which the sample of HVSs backpropagated in time most resemble a population of HVSs at the moment of ejection. A potential which cannot bring HVSs back to the GC  will be disfavored. 

We now turn back to $R(\textbf{w}_i)$, the distribution of possible initial configurations. When we generate our mock HVS population (see Sec. \ref{sec:methods:initial}), HVSs are initialized on the surface of a sphere 3 pc in radius with velocities pointed radially away from the GC.\footnote{This is not to imply that the HVS never existed at a $r<3$ pc -- rather, a main prediction of the \citet{Hills1988} is the velocity of the ejected star at an infinite distance from Sgr A*. In the Galaxy, this `infinite' distance can be approximated by $3$ pc, where Sgr A* becomes a subdominant contributor to the total Galactic potential. When the binary was disrupted and the HVS physically ejected at $r\ll3$ pc, the precise 3D velocity of the HVS depends on the geometry of the interaction \citep[see][]{Sari2010, Kobayashi2012, Rossi2014}. It is simpler in practice to impose an explicit ejection location at 3 pc with a strictly radial velocity.} In the formalism presented here, the initial configurations could therefore be expressed as
\begin{equation}
    R(\textbf{w}_i) = R_H(\abs{\textbf{v}_i}, m) \delta(\abs{\textbf{x}_i} - \mathrm{3 \; pc})\delta(\textbf{x}_i \cross \textbf{v}_i) \; \text{,}
\end{equation}
where the first delta function imposes the ejection distance of 3 pc from Sgr A* and the second ensures radial trajectories (zero angular momentum), and where $R_H(\abs{\textbf{v}},m)$ is the distribution of HVS masses and total velocities. $R_H(\abs{\textbf{v}_i}, m)$ is non-analytic in our HVS ejection scheme but can be approximated with a single power-law dependence on $m$ and a broken power-law dependence on $|\textbf{v}_i|$. We simplify things by saying $R_H(\abs{\textbf{v}_i}, m)\approx \text{const.}$, that is, we choose to remain agnostic towards the mass and total velocity distributions of HVSs ejected via the Hills mechanism.

In practice, $\mathcal{L}(\Theta)$ is evaluated by numerically propagating HVS orbits backwards in time. The presence of the Dirac deltas above mean $R(\textbf{w}_i)$ is a four-dimensional surface embedded in the seven-dimensional configuration space. Due to numerical effects it can never be satisfied exactly in application and the integrand will never evaluate as non-zero. Instead, we enlarge the ejection region by replacing the Dirac deltas in $R(\textbf{w}_i)$ with truncated Gaussians:

\begin{equation} \label{eq:R}
    R(\textbf{w}_i) \approx \exp(\frac{\abs{\textbf{x}_i} - \mathrm{3 \; pc}}{2\sigma_x}) \exp(\frac{\abs{\textbf{L}_i}}{2\sigma_L}) \; \text{,}
\end{equation}
where $\textbf{L}_i\equiv \textbf{x}_i\cross \textbf{v}_i$ is the angular momentum and $\sigma_x$ and $\sigma_L$ are smoothing parameters in both initial distance and angular momentum. They must be tuned to be a) small enough to effectively discriminate the better-fitting potentials, but b) large enough such that potentials slightly offset from the true potential may still have a non-zero likelihood. \citetalias{Contigiani2019} calibrate the method and suggest $\sigma_r = 10 \,\mathrm{pc}$ and $\sigma_L = 10\, \mathrm{pc \ km \ s^{-1}}$, because this is the order-of-magnitude angular momentum perturbation expected to be imposed on HVSs from two-body encounters with other stars. We truncate both Gaussians after four standard deviations. 

\citetalias{Contigiani2019} use a mock population of HVSs to show that this approach is effective at constraining the Galactic DM halo, though a degeneracy exists in the space of halo mass, radius and flattening. In Sec. \ref{sec:results} we ask whether this constraining power is hampered if the HVSs have traveled through a more complicated non-axisymmetric and time-varying potential. 

\vspace{18pt}
\subsubsection{The kinematic fitting method}\label{sec:methods:likelihood:gallo}

Rather than computing a trajectory for each HVS, the \citetalias{Gallo2022} approach opts instead to fit the HVS kinematics directly. Assuming an HVS is ejected from the GC on a radial trajectory, its initial tangential velocity in a spherical reference frame centered on the GC is zero. Along its trajectory, the star picks up a tangential velocity component $v_{\rm \theta}$ in the polar direction due to polar asymmetry in the potential, e.g. the Galactic disc or a flattened DM halo. If there are non-axisymmetric components of the potential, the star also attains a tangential velocity $v_{\rm \phi}$ in the azimuthal direction. Given a population of HVSs, one can compute $D_{\abs{v_{\theta}}}$, the distribution of the polar velocity magnitude, and also $D_{\tilde{v}_{\phi}}$, where $\tilde{v}_\phi \equiv v_\phi \text{sgn(tan}\phi)$ is the azimuthal velocity with a sign depending on the $\phi$ coordinate of the HVS. \citetalias{Gallo2022} show that the Galactic potential can be uniquely constrained by determining the potential which predicts $D_{\abs{v_{\theta}}}$ and $D_{\tilde{v}_{\phi}}$ distributions most similar to those of the observed HVS population.


Following \citetalias{Gallo2022}, we start by generating many \textit{mock HVS samples} propagated through several different MW-like potentials. Each potential consists of the same MBH, bulge and disc components as our MW-only potential as described in Sec. \ref{sec:methods:potential}. The DM halo of each potential has a scale mass and radius fixed at their respective fiducial values, but we vary the $y$-to-$x$ axis ratio $q_y$ and $z$-to-$x$ axis ratio $q_z$ on a grid, each spanning 0.5 to 1.3. For each ($q_y$, $q_z$) pair we initialize 100 samples of HVSs using the scheme described in Sec. \ref{sec:methods:initial} and propagate them through the potential. After selecting only stars with ejection velocities above 800 $\mathrm{km \ s^{-1}}$, positive present-day Galactocentric radial velocities and Galactocentric distances larger than 10 kpc, we are left with $\sim$900 stars per sample. 

For each sample we determine $D_{\abs{v_\theta}}$ and $D_{\tilde{v}_{\phi}}$. We then compare these distributions with those  
in our MW-only, MW+Bar, MW+LMC, MW+NIF and MW+All HVS populations, all of which were propagated under a potentials which include a $q_z=q_y=1.0$ DM halo. These comparisons are done using the two-dimensional two-sample Kolmogorov-Smirnov (KS) test with the \citet{Fasano1987} algorithm as implemented in the \textsc{python} package \textsc{ndtest}\footnote{\url{https://github.com/syrte/ndtest}}. This 
tests whether the null hypothesis (that the distributions are consistent with being drawn from the same parent distribution) is true. We reject this null hypothesis if the $p$-value of the test is smaller than significance level of 0.05. In other words, ideally the KS test $p$-values are consistently larger than 0.05 only for those mock HVS samples propagated through the DM halo with ($q_y$, $q_z$) identical to that through which our test populations were propagated. 

The \citetalias{Contigiani2019} approach and the \citetalias{Gallo2022} approach both offer formalisms for constraining the Galactic halo in different ways. A full accounting of the relative strengths and merits of each is beyond the scope of this work and we will treat them independently when we explore the degree to which non-axisymmetries and time dependence in the potential complicates them. Regardless, there are some differences to note. The \citetalias{Contigiani2019} approach is more interested in fitting the scale mass and size of the halo while the \citetalias{Gallo2022} approach is designed with constraining the halo shape in mind. We will stick to these respective use cases when we test them, though in principle each method can be used to constrain all parameters of the halo, or indeed all parameters of the total potential, simultaneously, although this would be a highly dimensional and degenerate parameter space. Notice that in the \citetalias{Contigiani2019} approach each HVS in an observed population contributes to $\mathcal{L}(\Theta)$ independently of the others. While a larger sample is preferred, this means that in principle a single HVS can rule out a disfavored potential, however this also makes the method susceptible to contaminant HVSs. The \citetalias{Gallo2022} approach is comparatively more robust to contamination since it fits using the entire population, but is most effective using a large population ($\sim$several hundred) of distant HVSs---the success rate of recovering the true potential decreases as the population size decreases. Such a population is not likely to be detected in the near-to-moderate future. 

It is important to note that both methods described above assume that both the positions and velocities of the HVSs are known to perfect accuracy. Since both would involve converting Earth-based observations to a Galactocentric rest frame, they each assume the position and velocity of the Sun in this frame are known to arbitrary precision as well. \citetalias{Gallo2022} defer an adjustment to the method to account for observational uncertainties to a future work, while the \citetalias{Contigiani2019} approach can in principle  handle observational uncertainties since the phase space position can be sampled over the errors and $\sigma_L$ and $\sigma_x$ (see Eq. \ref{eq:R}) can be scaled up if required. Regardless, in both methods very slight variations in HVS phase space coordinates can lead to significant changes in the best-fitting Galactic potential---obtaining such precise measurements in practice (e.g. measuring the tangential velocity of an HVS to within a fraction of a kilometer per second) is not feasible with current facilities.  

\section{Results}\label{sec:results}
In this Section we look at how the inclusion of the bar, LMC and its associated reflex motion impact HVSs in the MW+Bar, MW+LMC, MW+NIF, and MW+All potentials when compared to the MW-only potential. First, we examine how closely the HVSs trace back to the GC in each of the above potentials. We then apply the fitting methods described in Sec. \ref{sec:methods-likelihood}. Next, we examine how HVSs deviate from radial trajectories. Lastly, we investigate whether Gaia DR4 predicted performance would be sufficient to measure these effects. 

\subsection{HVS Trajectories}\label{sec:results-traj}

\begin{figure*}
    \centering
    \includegraphics[width=\linewidth]{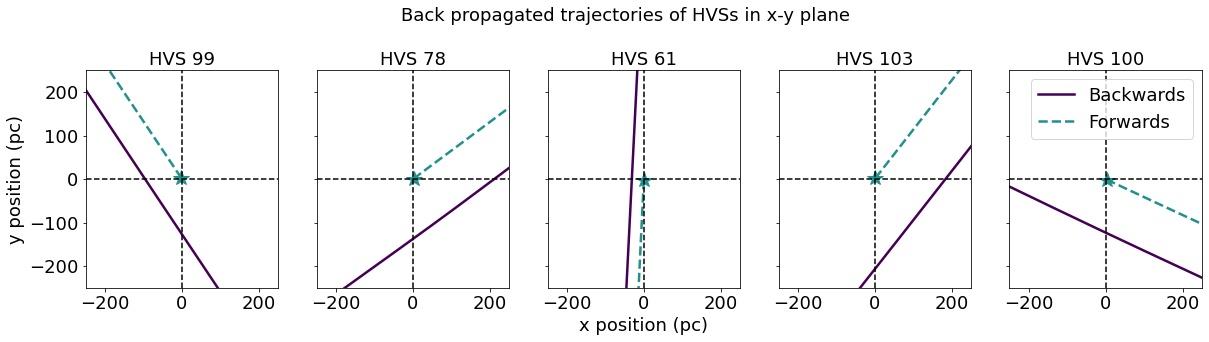}
    \includegraphics[width=\linewidth]{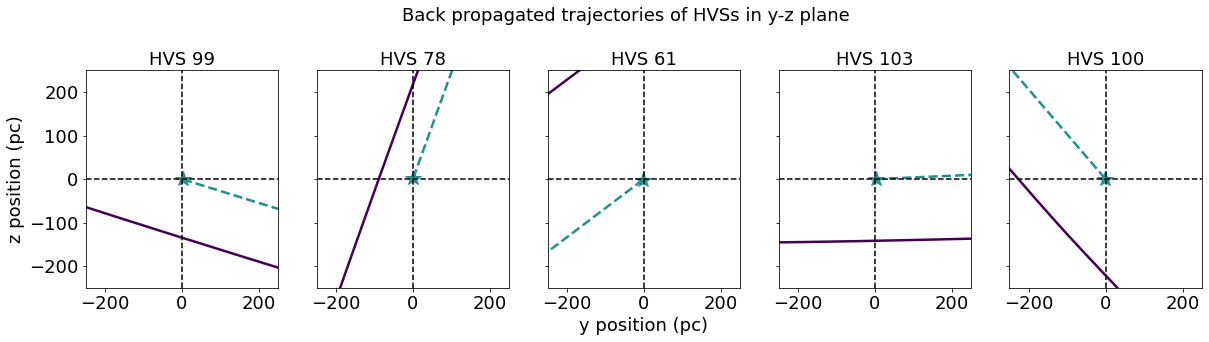}
    \caption{Trajectories of representative HVSs in the Galactocentric $x-y$ and $y-z$ planes. Forward trajectories (blue) are in our MW+All potential, backwards trajectories (orange) are in our MW-only potential.
    }
    \label{fig:Obstraj}
\end{figure*}

\begin{figure}
    \centering
    \includegraphics[width=\columnwidth]{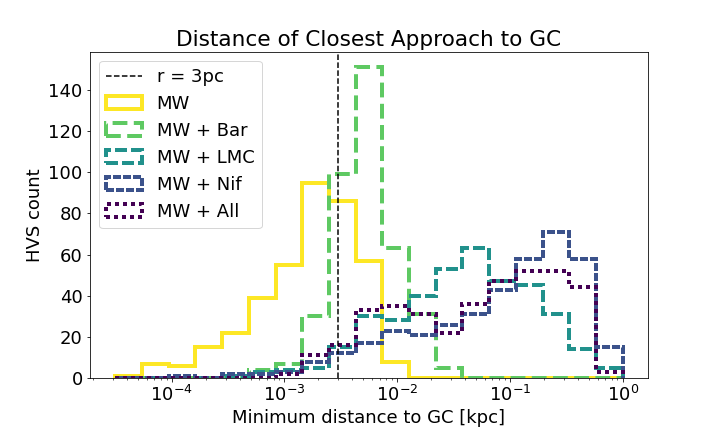}
    \caption{Histogram showing how close HVS propagated forward in a given potential gets to the GC if they are propagated backwards in time in the fiducial MW-only potential.}
    \label{fig:BarLMCHist}
\end{figure}

\begin{figure}
    \centering
    \includegraphics[width=\columnwidth]{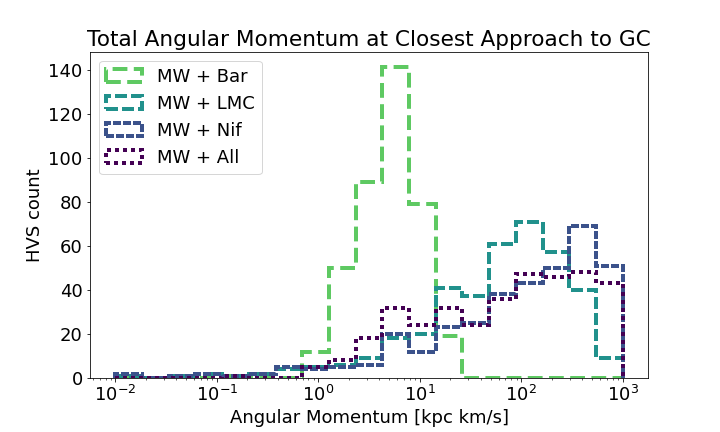}
    \caption{Histogram showing the total angular momentum $L$ at closest approach to the GC for stars propagated outwards in a given potential and backwards in the MW-only potential. HVSs propagated in the MW potential return with near zero angular momentum, while HVSs in potentials containing non-axisymmetric and time-dependent components have picked up significant angular momentum during back-propagation. The angular momentum for the MW sample sits at $10^{-7}$, and has not been included in order to highlight the non equilibrium behavior.}
    \label{fig:BarLMCAngMom}
\end{figure}

To gain some intuition for how the impact of the bar, LMC, and reflex motion impact HVS kinematics, we begin by demonstrating explicitly how these components affect the trajectories of HVSs. We propagate a sample of HVSs through our MW+All potential. Then we propagate the sample backwards in time through our MW-only potential to find where they would appear to originate under the assumption that the bar and LMC contributions to the Galactic potential are negligible. \par 

Fig. \ref{fig:Obstraj} displays the forward and backward trajectories in the Galactocentric $x-y$ and $y-z$ planes of five representative HVS. We find that all these stars regardless of ejection direction will appear to miss the GC by between 68 and 328 pc if propagated backwards through a simplistic potential. These offsets may seem small on the scale of the whole Galaxy, but changes in apparent ejection location of this order can make the difference between groundbreaking HVS discoveries such as S5-HVS1 \citep{Koposov2020} and HVS candidates such as Gaia DR3  3126801097033888768 \citep{Marchetti2022} and DR3 4094201527955913856 \citep{Liao2023} which are interesting but should nonetheless viewed with skepticism. 

Rather than showing a few representative stars, in Fig. \ref{fig:BarLMCHist} we explore the apparent change in ejection location for the whole sample. Here we show the distributions of closest approach distances to the GC for stars propagated forwards through each of our test potentials and backwards in the MW-only potential. Here we use smaller test samples propagated with a timestep of 0.005 Myr (10x smaller than that used in the main samples) to mitigate the impact of numerical effects. We find that neglecting the bar, LMC or reflex motion on back-propagation can lead individual HVSs to miss the GC by up to several hundred parsecs. Stars propagated forwards through MW+All miss the GC by the greatest amount, followed by MW+LMC, MW+NIF and lastly MW+Bar. Additionally, the inclusion of non-axisymmetric components means that when stars are back-propagated in the MW-only potential, they reach their closest approach to the GC with significant angular momentum (Fig. \ref{fig:BarLMCAngMom}). In the MW-only potential, the angular momentum of the backpropagated HVSs does not exceed $L=10^{-6}$ kpc km s$^{-1}$. In contrast, HVSs in the MW+All population have a minimum of $L = 0.016$ kpc km s$^{-1}$ and a maximum of $L = 2415$ kpc km s$^{-1}$. In the case of the MW-only potential, the HVSs do not miss the GC and have zero angular momentum, and so we can rule out numerical issues as the source of this discrepancy and instead attribute it to the combined influence of bar, LMC and reflex motion. \par 

\subsection{Potential Reconstruction} \label{sec:results-potential}

\begin{figure*}
    \centering
    \includegraphics[width=1.06\columnwidth]{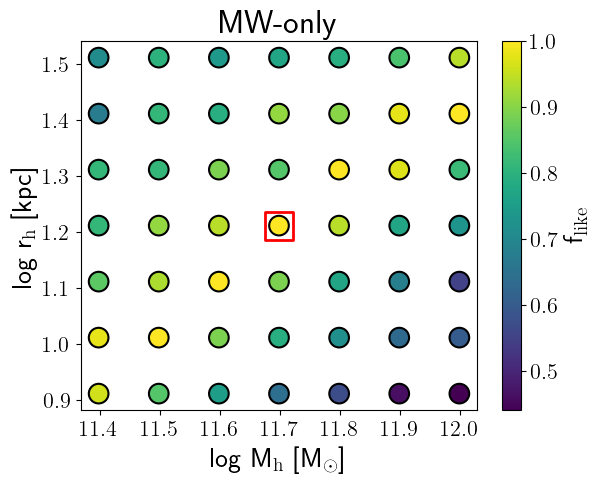}
    \includegraphics[width=1.05\columnwidth]{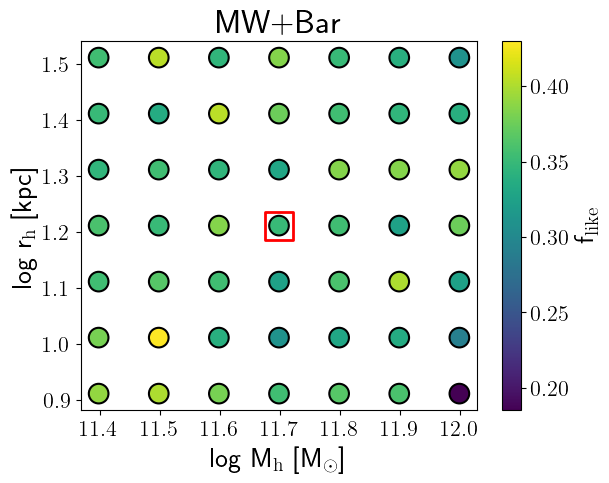}
    \includegraphics[width=1.05\columnwidth]{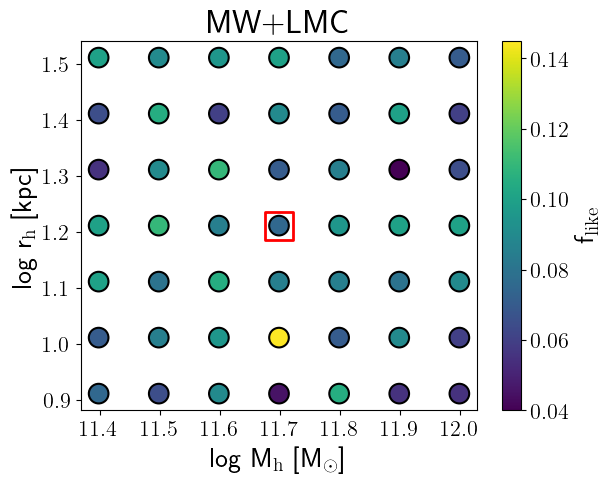}
    \includegraphics[width=1.05\columnwidth]{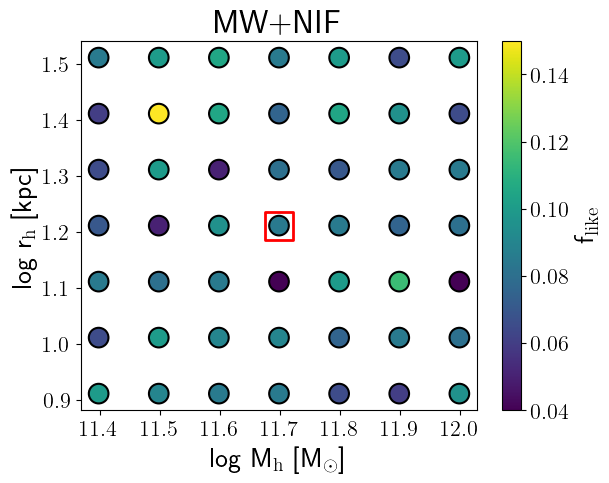}
    \includegraphics[width=1.05\columnwidth]{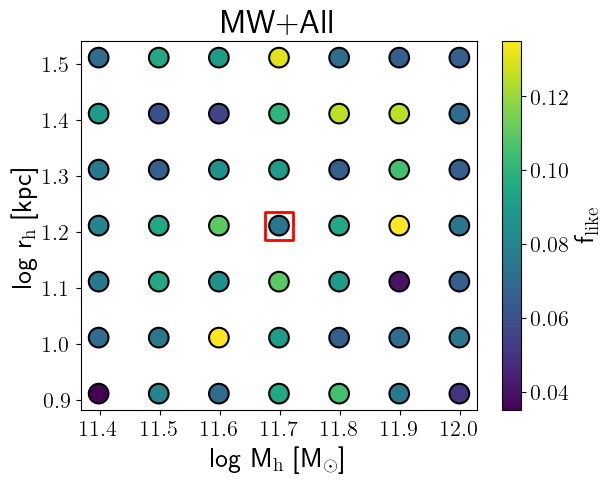}
    \caption{The fraction of HVSs among the MW-only, MW+Bar, MW+LMC, MW+NIF and MW+All samples which contribute a non-zero likelihood for different potentials including a spherical NFW halo of varying scale mass and scale radius, following the formalism of \citetalias{Contigiani2019}. See text for details. The red square marks our fiducial halo parameters of $M_{\rm h} =0.76\times10^{12} \; \mathrm{M_\odot}$ and $r_{\rm h} = 24.8$ kpc. Notice that the colorbar is not fixed across all panels.}
    \label{fig:likelihood}
\end{figure*}

\begin{figure}
    \centering
    \includegraphics[width=0.9\columnwidth]{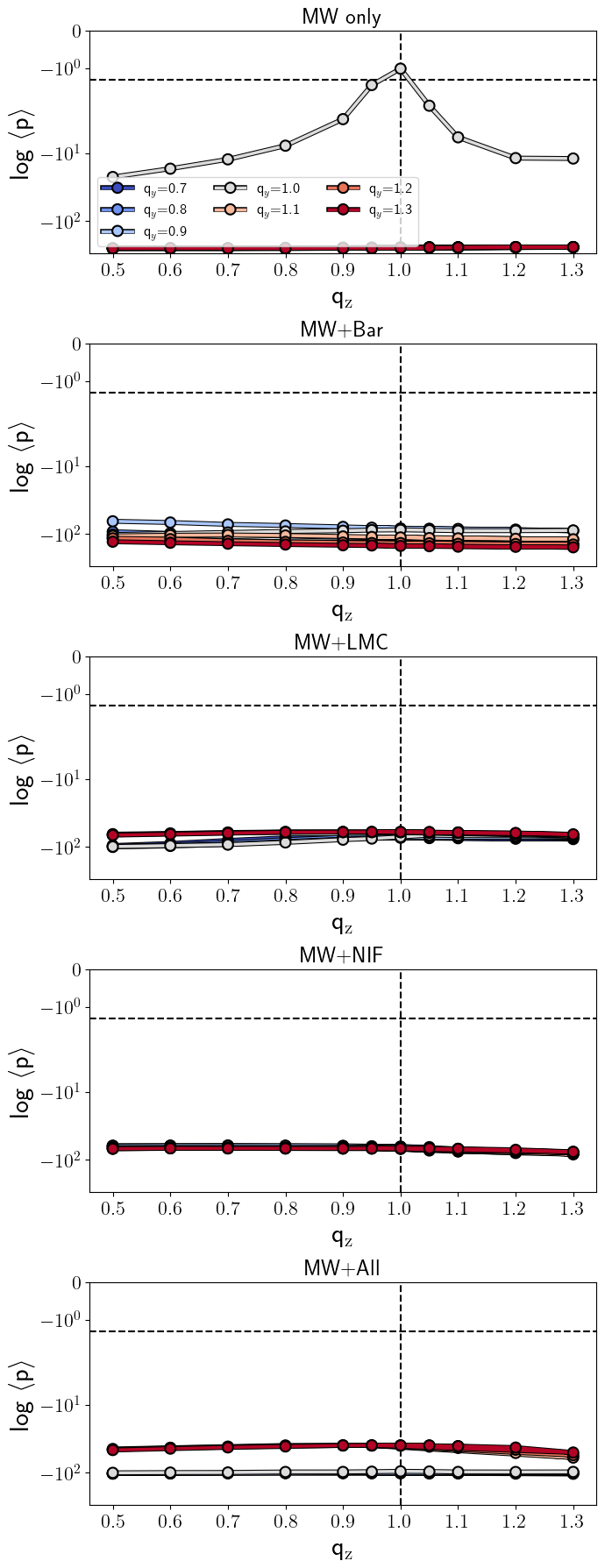}
    \caption{The median $p$-values of the two-dimensional KS tests used in the \citetalias{Gallo2022} formalism for constraining the Galactic DM halo. See text for details. Median $p$-value against DM halo axis $z$-to-$x$ ratio $q_z$ are shown for different values of the $y$-to-$x$ axis ratio $q_y$ (different lines) for test HVS samples propagated through different model (different panels).}
    \label{fig:Gallo2D}
\end{figure}

We now turn to the Milky Way gravitational potential recovery approaches outlined in Sec. \ref{sec:methods-likelihood} to investigate the degree to which they are complicated by the consideration of non-axisymmetric and time-dependent potential components. We start with the \citetalias{Contigiani2019} formalism and our approach is as follows: we take our MW sample of HVSs and propagate them backwards in time through a set of potentials with identical MBHs, bulges and discs as our default MW potential but with DM halos varying in a grid spanning scale masses from $M_{\rm h} = 10^{11.4} \; \mathrm{M_\odot}$ to $M_s = 10^{12.0} \; \mathrm{M_\odot}$ and scale radii spanning 8 kpc to 32 kpc. As a sanity check, this grid includes the DM halo corresponding to our fiducial MW potential with $r_{\rm h} = 16.244$ kpc and $M_{\rm h} =0.5 \times 10^{12} \; \mathrm{M_\odot}$. We integrate these trajectories through Eq. \ref{eq:R} to evaluate $\mathcal{L}_j$ (Eq. \ref{eq:L}) for each star in each potential. We then do the same for all stars in our MW+Bar, MW+LMC, MW+NIF and MW+All samples, integrating them back in time through the same grid of axisymmetric Milky Way potentials. For this calculation we keep $\sigma_{\rm r}=10$ pc but increase $\sigma_{\rm L}$ by a factor of 100 from the value suggested by \citetalias{Contigiani2019} of $\sigma_{\rm L}=10 \; \mathrm{pc \ km \ s^{-1}}$ to $\sigma_{\rm L}=1000 \; \mathrm{pc \ km \ s^{-1}}$ since we are working with backwards trajectories which approach the GC with greater angular momentum (see Fig. \ref{fig:BarLMCAngMom}). 

The results of this calculation are shown in Fig. \ref{fig:likelihood}. Rather than showing $\mathcal{L}=\sum_j \mathcal{L}_j$, here we plot the fraction $f_{\rm like}$ of HVSs from each sample which have a non-zero $\mathcal{L}_j$ for different Milky Way DM halos of varying mass and size. This fraction is more informative in this case than the numerical $\mathcal{L}$ since it is less susceptible to stochastic variation and numerical effects from the orbital integration. For the MW-only sample (top), 100\% of stars have a non-zero $\mathcal{L}_j$ in our fiducial DM halo (red square), as expected. We recover the degeneracy in this space found by \citetalias{Contigiani2019}, in which halos at fixed $M_{\rm h}/r_{\rm h}^2$ are equally favored, since the HVSs are sensitive to the gravitational force from the halo rather than its mass or size directly. For halos far from this degeneracy line, $f_{\rm like}$ can fall to as low as 45\%. 

Turning to the remaining potentials, for the sample propagated through the MW+Bar potential (middle left), our fiducial DM halo no longer has the largest $f_{\rm like}$. In fact, no halo in this space has an $f_{\rm like}$ exceeding 50\%, nor are there any halos in this space which can be strongly disfavored relative to others. If the parameters of the DM halo through which these stars were originally propagated were not known a priori, it would be impossible to recover them from this sample -- it appears as if the presence of the bar has ruined the ability of this approach to constrain the halo. The same can be said of the other samples, particularly for the MW+LMC and MW+All potentials, where no more than 15\% of HVSs give a non-zero $\mathcal{L}_j$ for any potential. These plots show decisively that when the potential that has influenced their trajectories includes non-axisymmetric or time-dependent elements, even relatively minor ones, HVSs lose all their power to constrain a static, axisymmetric Milky Way potential. Even if $\sigma_r$ and/or $\sigma_L$ are made larger or smaller, this fact remains true. 

We next turn to how the bar and LMC impact the \citetalias{Gallo2022} approach to constraining the halo as outlined in Sec. \ref{sec:methods:likelihood:gallo}. In Fig. \ref{fig:Gallo2D} we show the median $p$-values resulting from the two-dimensional KS test of the distributions of $v_\theta$ and $\tilde{v}_\phi$ from our MW-only and MW+Bar/LMC/NIF/All test samples against those of mock samples propagated through DM halos of varying shape. Notice that the log $p$-value is itself shown in log scale---the smallest $p$-values on this plot approach $10^{-300}$. For the MW-only test sample propagated through a spherical ($q_z=q_y=1.0$) halo, we see that indeed only KS tests against mock samples with $q_y=1.0$ have $p$-values above $10^{-10}$ and of these, only the tests against the mock samples propagated through the halos with  $q_z$ very near to 1.0 yield a $p$-value above our acceptance threshold of 0.05. For all of the remaining test samples propagated through non-axisymmetric and time-dependent potentials, there is no combination of $q_z$ and $q_y$ whose mock samples yield a $p$-value even approaching this threshold within ten orders of magnitude. Without prior knowledge of the DM halo through which these test samples were propagated, this approach cannot provide any insight. We note that for $p \ll 0.05$, the numerical value of $p$ is not particularly meaningful. Choices in the specific implementations of the 2D KS test algorithm and computational effects can lead to changes in $p$ of orders of magnitude.

Together, the results from both formalisms demonstrate that it is crucial to consider minor components of a Milky Way potential model, especially those which vary in time, when attempting to use HVSs as a tracer for the Galactic potential. While these components are subdominant in terms of the total mass of the Galaxy, their subtle impacts on HVS trajectories greatly impact the feasibility of reconstructing the Galactic potential. 

\subsection{HVS Deflection}\label{sec:results-deflect}

We have so far explored the implications of the bar, LMC, and reflex motion on HVS trajectories and on the prospect of reconstructing the Galactic potential using HVSs. However, these are directly observable. With observatories and missions such as Gaia, stellar positions, velocities, and distances can be measured. To explore explicitly how the bar, LMC, and reflex motion impact HVS kinematics, we define the angle $\beta$: the angle between a HVS's position vector in Galactocentric coordinates and its velocity vector. Due to their extreme velocities and the relatively short time they spend deep within the potential of the Galaxy, $\beta$ is typically assumed to be small and $\beta\approx0$ is commonly used as a discriminant to identify HVS candidates \citep{Marchetti2017, Bromley2018, Verberne2024, Scholz2024}. If non-equilibrium components of the Galactic potential contribute to non-zero $\beta$, it could bias the populations of HVSs uncovered in the future. 

We split $\beta$ into the azimuthal component $\beta_\phi$, the difference between the HVSs position and velocity vector in the plane of the disc:
\begin{equation}
    \beta_\phi = \text{sgn} ( \textbf{v}_R \times \textbf{R})\cdot \text{acos} \left( \frac{\textbf{v}_R \cdot \textbf{R}}{|\textbf{v}_{R} ||\textbf{R}|} \right) = \text{atan2} ( \textbf{v}_R \times \textbf{R},\textbf{v}_R \cdot \textbf{R}) \; ,
\end{equation}
where $\textbf{R}=(x,y)$ and $\textbf{v}_{R}=(v_x,v_y)$ are the kinematics in the plane of the disc. We similarly define the  component $\beta_\theta$, the angular deflection in the polar direction;
\begin{equation}
    \beta_\theta = \text{sgn} ( \textbf{v} \times \textbf{r})\cdot \text{acos} \left( \frac{\textbf{v} \cdot \textbf{r}}{|\textbf{v} || \textbf{r}|} \right) \; ,
\end{equation}
where $\textbf{r}=(\sqrt{x^2+y^2},z)$ and $\textbf{v}=(\sqrt{v_x^2 + v_y^2},v_z)$ are the kinematics in cylindrical coordinates. To clarify, we use the ISO convention where the inclination angle $\theta$ is measured from the positive $z$-axis downwards. Note that the definitions used in \citet{Boubert2020} are yet another equivalent expression;
\begin{align}
    \beta_\phi &= \text{atan2} (y,x) - \text{atan2} (v_y,v_x) \\
    \beta_\theta &= \text{acos}(z/|\textbf{r}|) - \text{acos} (v_z/|\textbf{v}|) \; .
\end{align}

The signs of $\beta_\theta$ and $\beta_\phi$ indicate the direction in which an HVS is being deflected off of a radial trajectory. Positive $\beta_\phi$ indicates counter clockwise deflection in the $x-y$ plane and negative values indicate clockwise deflection, and similarly positive $\beta_\theta$ indicates deflection towards increasing $\hat{\theta}$ (i.e. towards the $-z$ direction). In our axisymmetric MW-only potential, we expect $\beta_\phi$ to be zero everywhere, and indeed in our simulations we find $|\beta_\phi|\simeq10^{-8}$ rad, which aligns with expectations plus some numerical noise. The presence of the Galactic disc introduces non-zero $\beta_\theta$, which we find to be up to $\pm 10^{-2}$ rad in our simulations. For our non-axisymmetric potentials we additionally define $\Delta \beta_\theta \equiv \beta_\theta - \beta_{\theta,MW}$, the difference between $\beta_\theta$ of a star propagated in the potential of interest compared to that same star propagated in our MW-only potential. This allows for the residual effects of the non-axisymmetric components to be investigated directly\footnote{When comparing the kinematics of single stars propagated through different potentials, there are edge cases in which a star is barely unbound in our MW-only potential but marginally bound to the Galaxy in a non-equilibrium potential or (even more rarely) vice versa. Though they may not technically qualify as genuine HVSs depending on the chosen potential, we do not remove these stars from our sample since they do not affect the overall results presented here.}.

\begin{figure*}
    \centering
    \includegraphics[width=\linewidth]{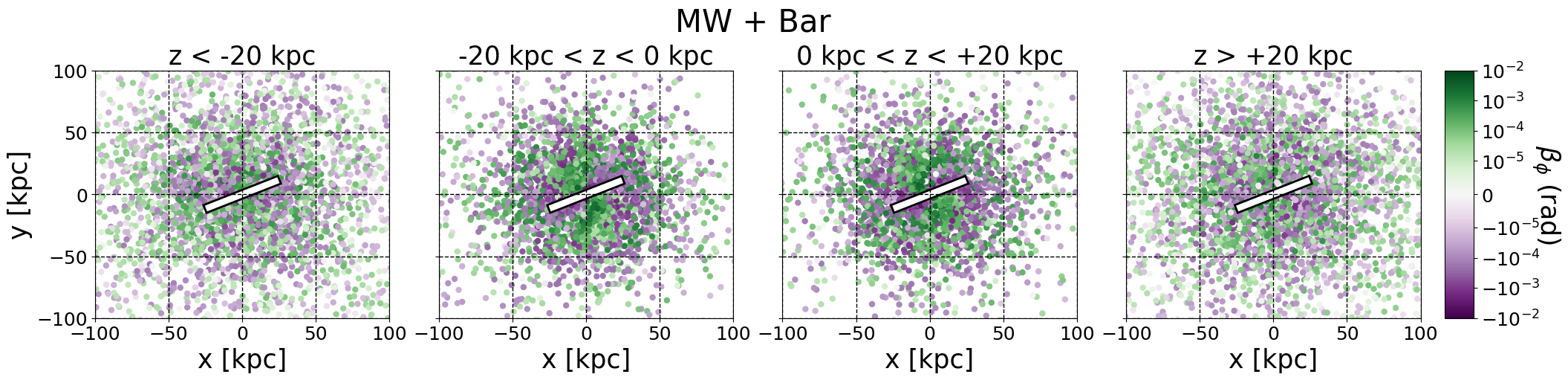}
    \includegraphics[width=\linewidth]{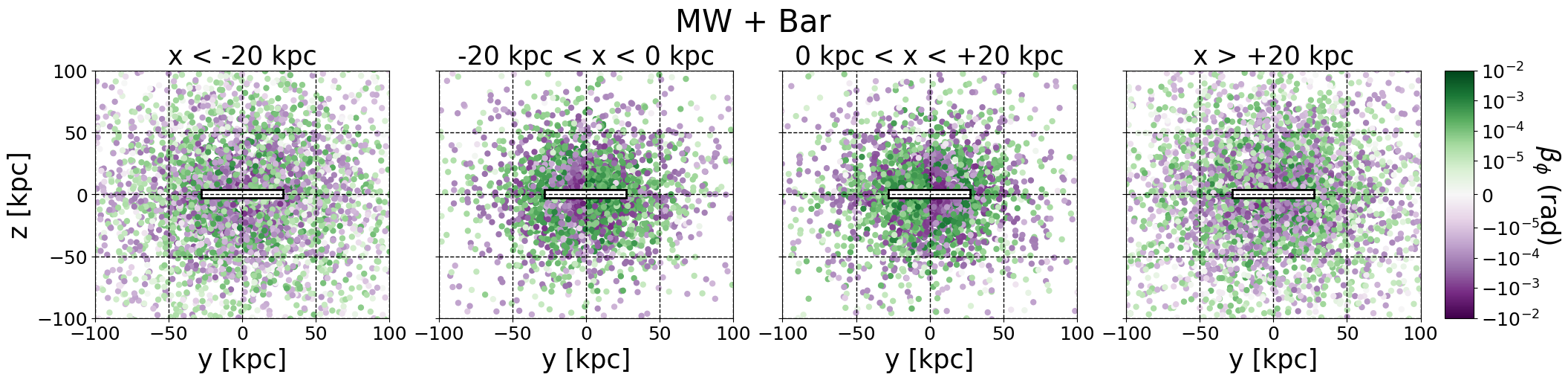}
    \includegraphics[width=\linewidth]{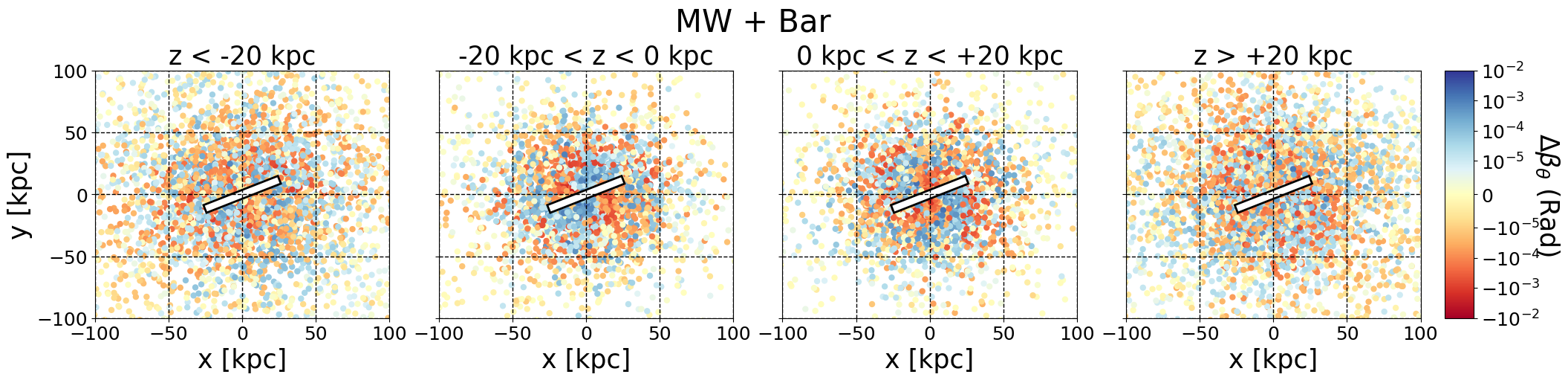}
    \includegraphics[width=\linewidth]{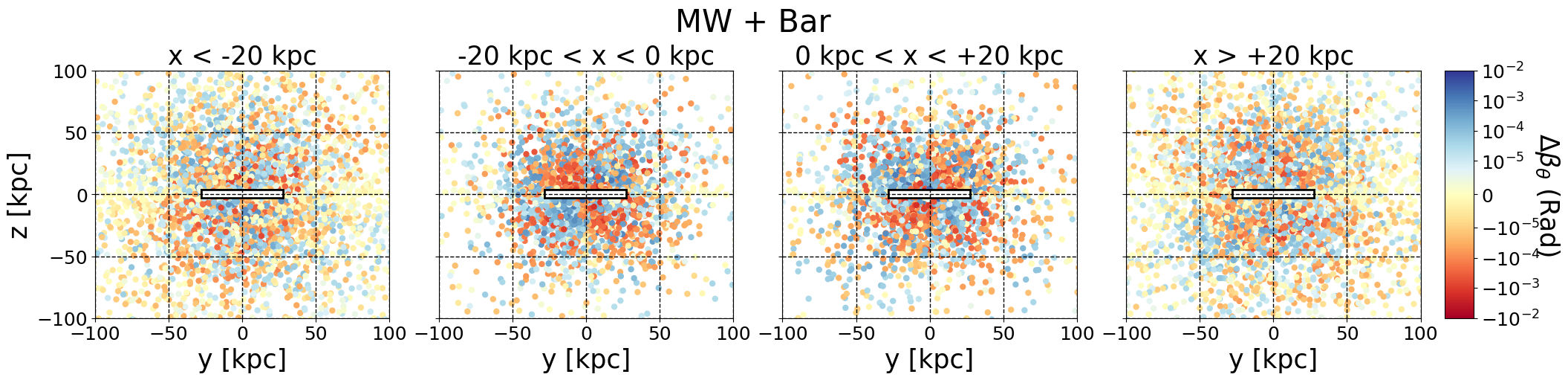}
    \caption{Impact of the Galactic bar on HVS trajectories. The bar location is shown in white, and each dot represents an individual HVS. Each HVS is colored by the deflection imparted on it by the bar. The top two panels plot $\beta_\phi$ in the $x-y$ (1st row) and $y-z$ (2nd row) planes. Positive HVSs are deflected counter-clockwise, and negative HVSs are deflected clockwise. The bottom two rows plot $\Delta\beta_\theta$, which is the excess $\beta_\theta$ in the MW+Bar potential compared to the same star in the MW potential. Negative color indicates HVS are deflected more north than in the MW case, and positive indicates more to the south than the MW case. Each row contains four panels which slice the HVS sample by distance in either the z of x direction, depending on the projection. The white bar shows the orientation of the Galactic bar in the appropriate projection, though the length is not to scale.}
    \label{fig:beta_bar}
\end{figure*}

\begin{figure*}
    \centering
    \includegraphics[width=\linewidth]{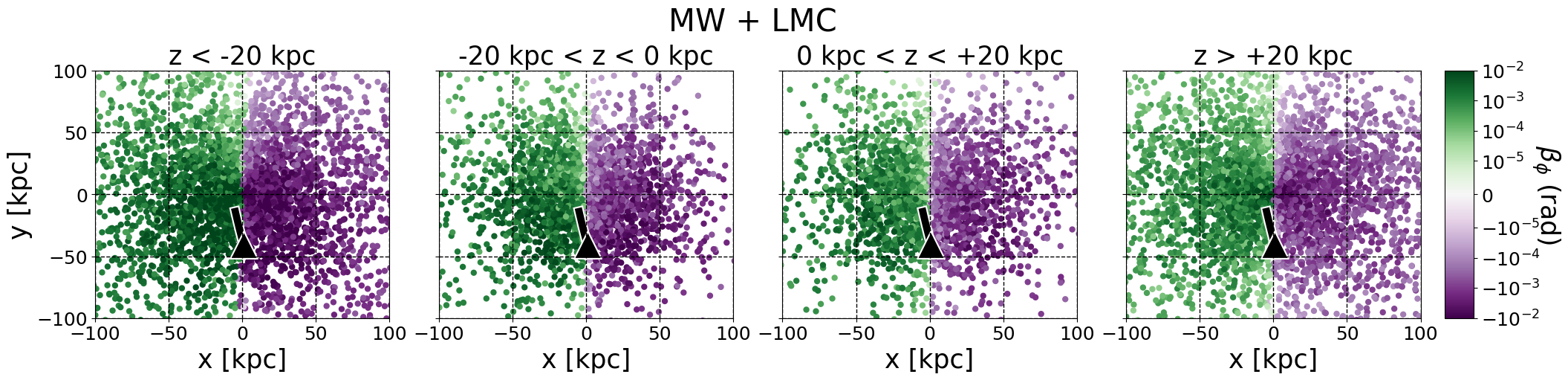}
    \includegraphics[width=\linewidth]{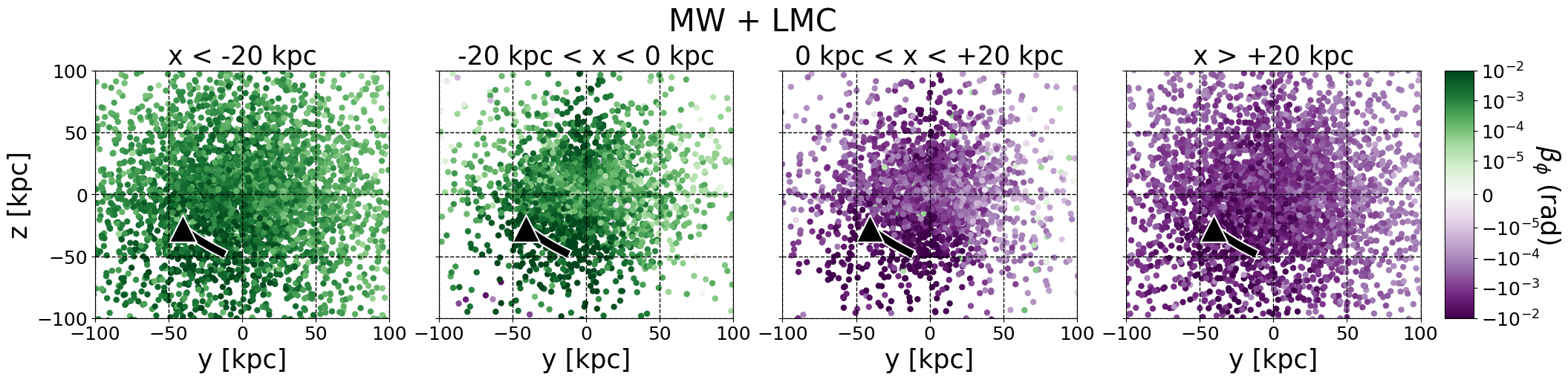}
    \includegraphics[width=\linewidth]{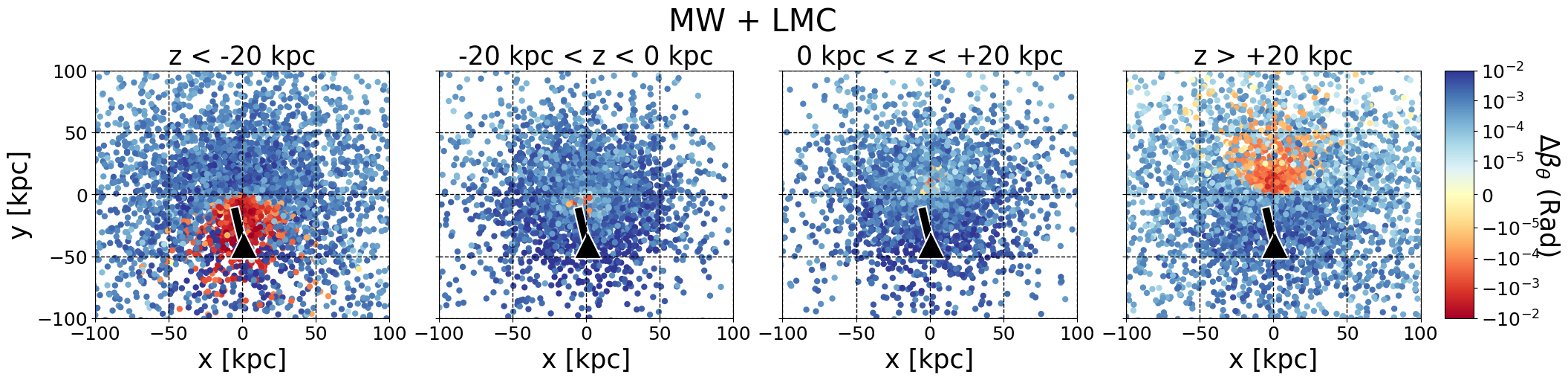}
    \includegraphics[width=\linewidth]{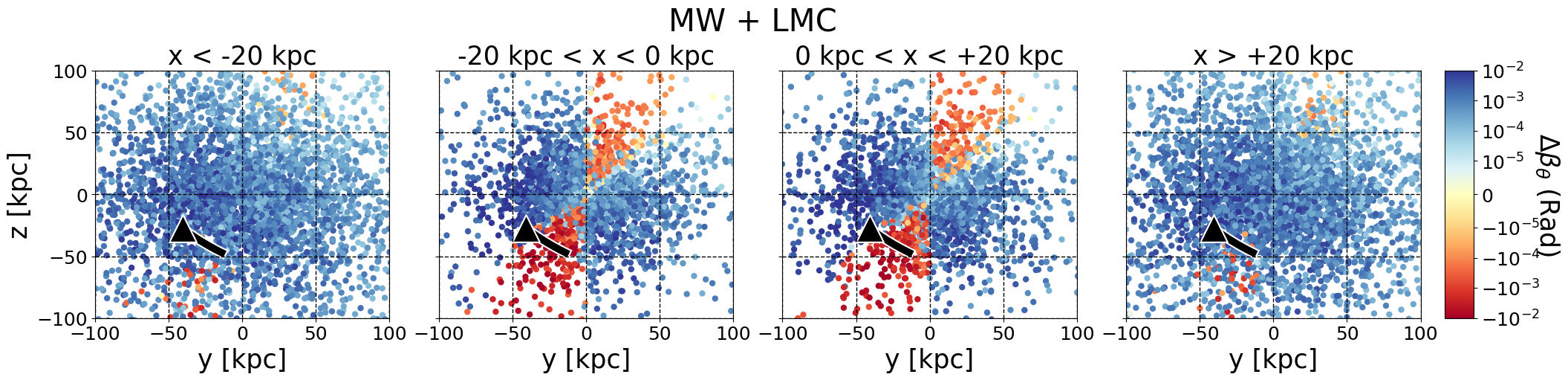}
    \caption{Similar to Fig. \ref{fig:beta_bar} but showing the impact of the LMC on HVS trajectories. The location of the LMC is indicated by the black triangle, and the black line shows its orbit over the last 100 Myr.} 
    \label{fig:beta_LMC}
\end{figure*}

\begin{figure*}
    \centering
    \includegraphics[width=\linewidth]{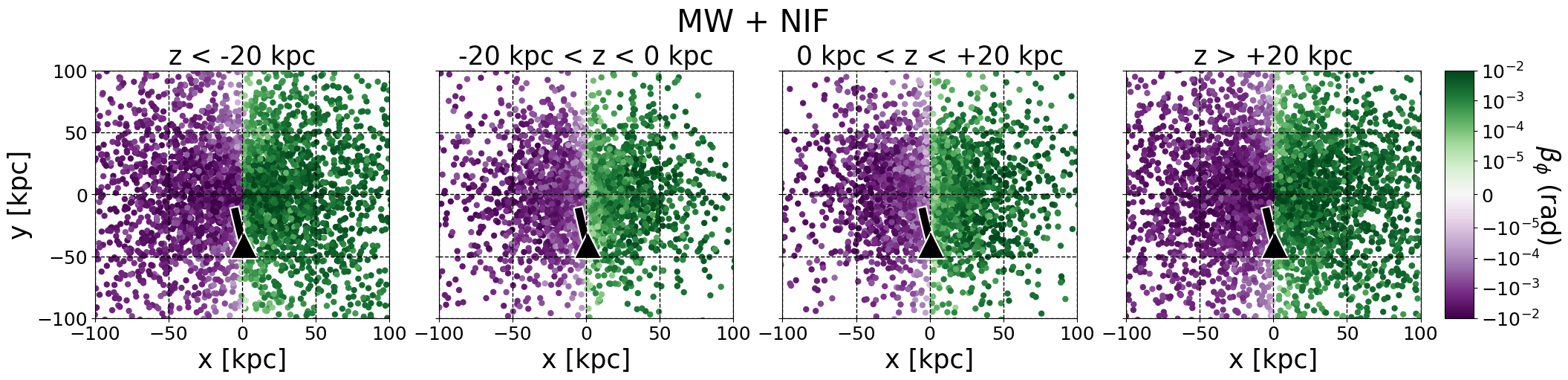}
    \includegraphics[width=\linewidth]{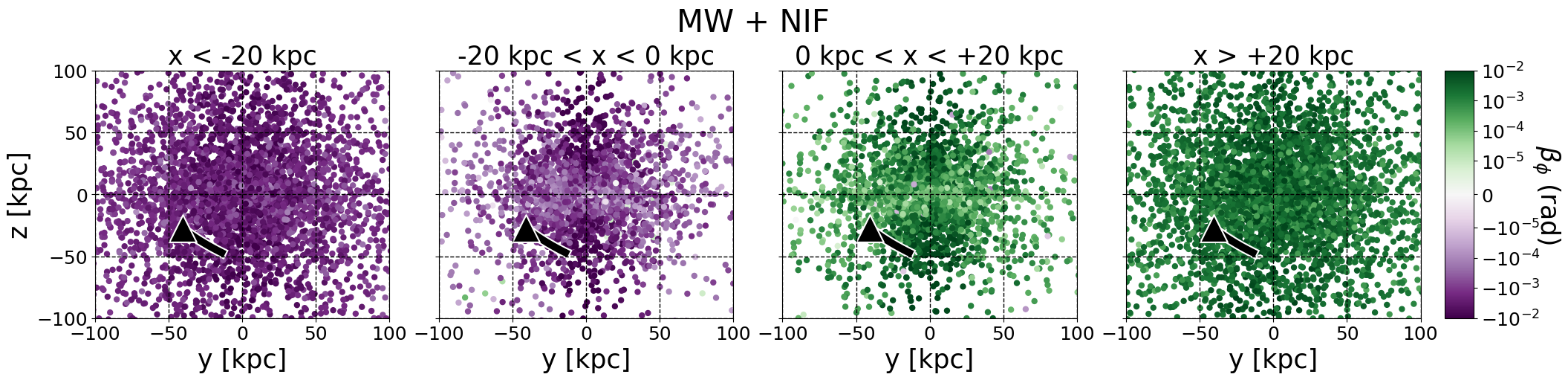}
    \includegraphics[width=\linewidth]{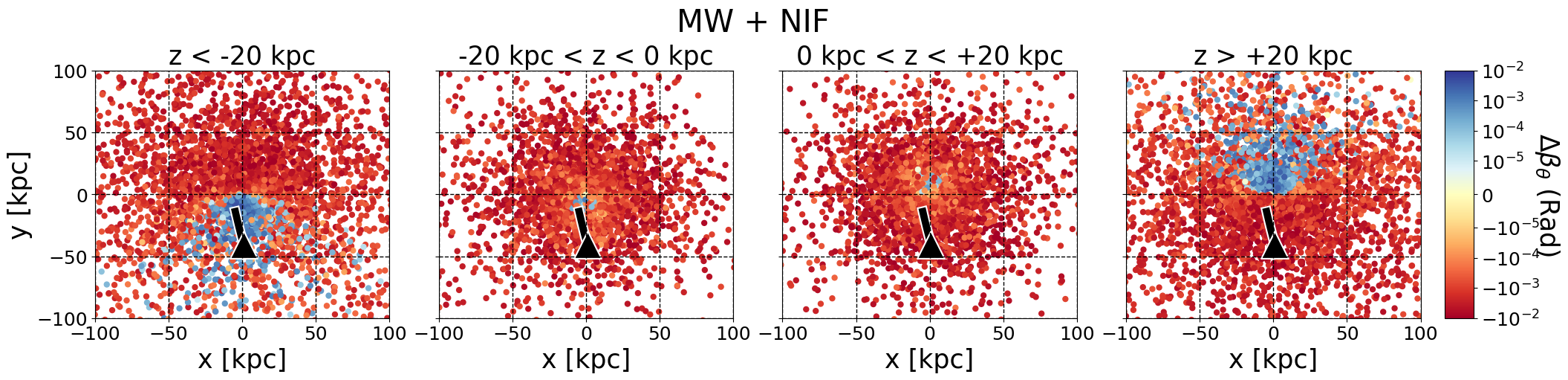}
    \includegraphics[width=\linewidth]{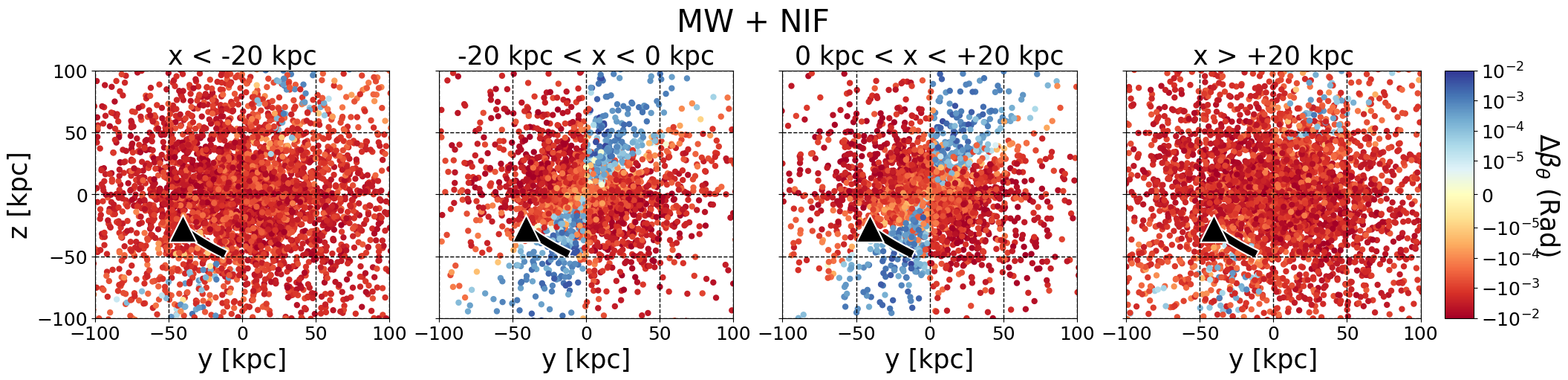}
    \caption{Similar to fig. \ref{fig:beta_LMC} but showing the impact of the non-inertial reference frame crated by the LMC on HVS trajectories. }
    \label{fig:beta_nif}
\end{figure*}

\begin{figure*}
    \centering
    \includegraphics[width=\linewidth]{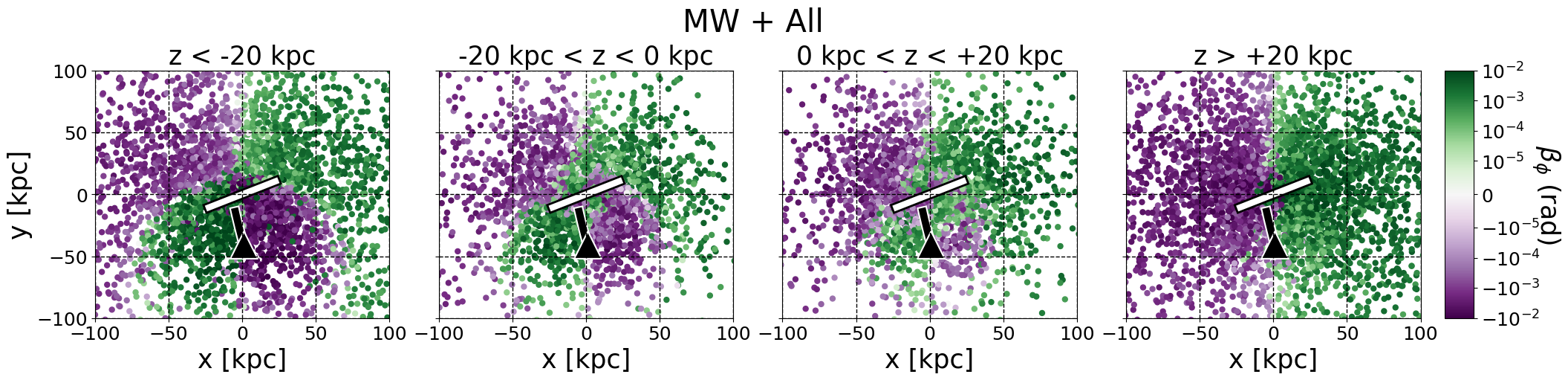}
    \includegraphics[width=\linewidth]{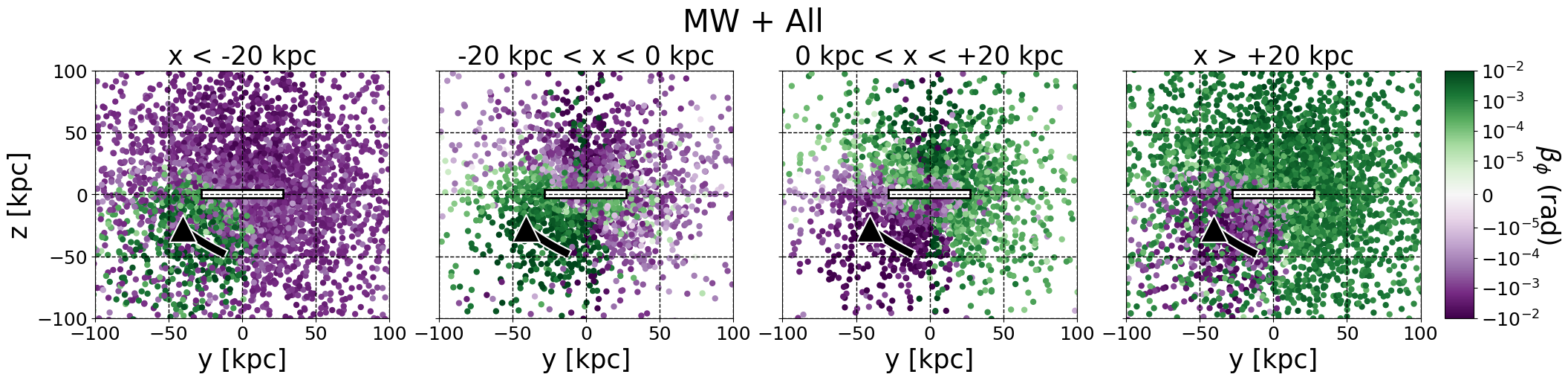}
    \includegraphics[width=\linewidth]{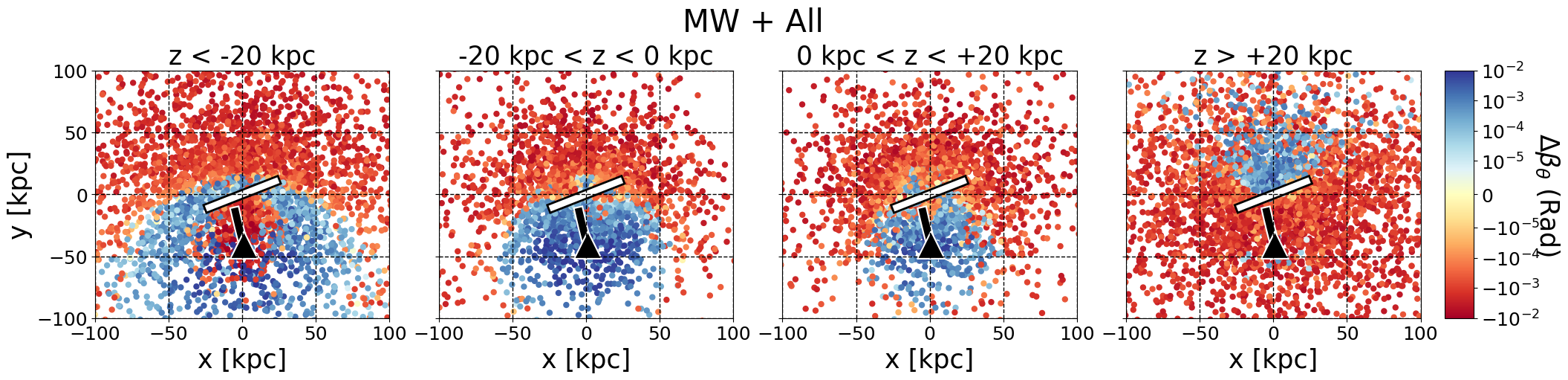}
    \includegraphics[width=\linewidth]{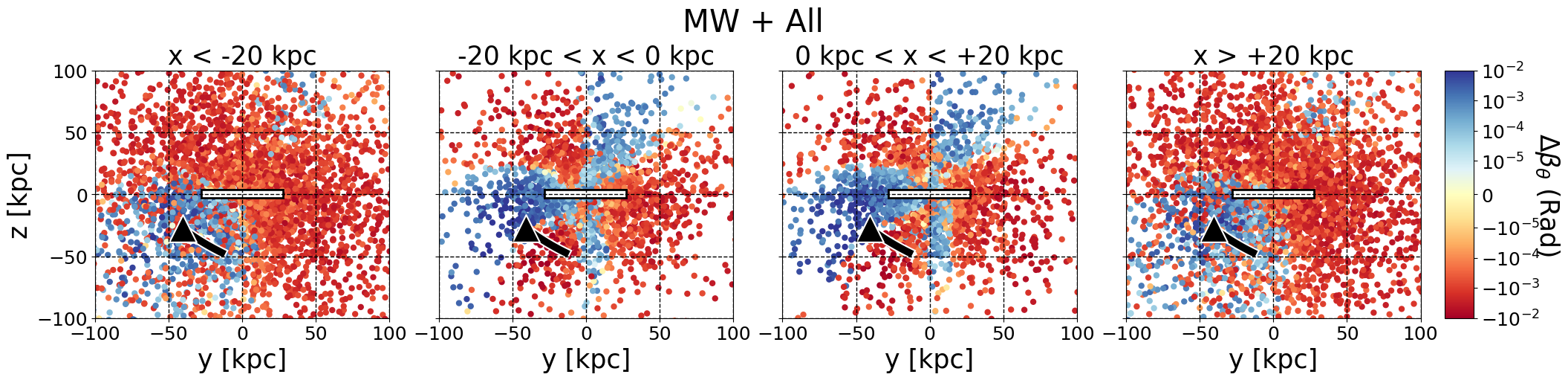}
    \caption{Similar to Fig. \ref{fig:beta_bar} but showing the impact of the MW+All potential on HVS trajectories.}
    \label{fig:beta_all}
\end{figure*}

Figs. \ref{fig:beta_bar} to \ref{fig:beta_all} plot the deflection of HVSs in our four non-axisymmetric, time-dependent potentials of interest. The first two rows plot $\beta_\phi$, with the first row plotting $\beta_\phi$ in the $x-y$ plane and the second row plotting $\beta_\phi$ in the $y-z$ plane. The third and fourth rows plot $\Delta \beta_\theta$ in the $x-y$ (third row) and $y-z$ (fourth row) projections. The columns split the HVS sample by the third dimension, chosen to be $z< -20$ kpc, $-20$ kpc $< z < 0$ kpc, $0$ kpc $ < z < +20$ kpc, and $z >+20$ kpc in the $x-y$ projection, and similarly with $x$ for the $y-z$ projections. In each panel, individual dots represent single HVSs, and are colored based on the deflection of that particular HVS. \par 

Fig. \ref{fig:beta_bar} plots $\beta_\phi$ and $\Delta\beta_\theta$ for the MW+Bar potential. Some stars are deflected clockwise while others are deflected counter-clockwise depending on where they are in relation to the bar axis, and likewise some stars are deflected more toward the Galactic midplane than they otherwise would be in the MW=only potential. The spiral structure close to the GC in the $x-y$ projection for both $\beta_\phi$ and $\beta_\theta$ is due to the rotation of the bar. Stars ejected parallel or perpendicular to the bar will experience minimal $\beta_\phi$ as there is net $\sim$zero azimuthal torque along their initial trajectories, while $|\beta_\phi|$ is maximized for stars ejected with azimuthal angles at odd intervals of $\pi$/4 relative to the bar axis. If the bar were stationary, these critical angles would be fixed in the $x-y$ plane. However, since the bar rotates, the bar axis at the moment of ejection depends on the flight time of the star and the alternating positive and negative $\beta_\phi$ behavior winds in a spiral pattern. For reference, for an HVS at $30$ kpc from the GC where the winding pattern in $\beta_\phi$ becomes apparent, the bar has completed anywhere from 5\% to 35\% of a full rotation between the initial ejection time and the present day depending on the ejection velocity of the star. In the $y-z$ projection, $\beta_\phi$ is mirrored over the x-axis, and shows further that the azimuthal deflection is strongest for stars nearer to the GC. Similar spiral patterns appear in $\Delta \beta_\theta$ in the $x-y$ projection, and can be explained with similar logic---stars ejected along the bar axis will be deflected more than in the MW-only potential since there is more mass concentrated at the Galactic midplane locally, and stars ejected perpendicular to the bar are deflected less than in the MW-only case. $\Delta \beta_\theta$ flips in sign above and below the disc depending on whether $\hat{\theta}$ is directed toward or away from the Galactic midplane. 

Fig. \ref{fig:beta_LMC} is similar to Fig. \ref{fig:beta_bar} but for the MW+LMC potential. Here, the symmetry is different because the LMC is located below the plane of the disc at $x \approx +1$ kpc, $y\approx-40.6$ kpc, $z\approx -27.5$ kpc ($\phi\approx=-90^\circ, \theta=214^\circ$). In $\beta_\phi$ we see a clear division between HVSs deflected clockwise vs counter-clockwise in the $\beta_\phi$ plots. HVSs located at $x> 0$ are all deflected clockwise, whereas the HVS located at $x<0$ are deflected counter-clockwise. In both cases, HVS are deflected towards the LMC. HVS located below the disc are deflected more compared to HVS located above the disc, since they are physically closer to the LMC and therefore more susceptible to its influence. In the $x-y$ plane, stars at $z < -20$ kpc experience both northward and southward $\Delta\beta_\theta$. This is because some HVS in this slice are located above the LMC while others are located below it. In either case, the HVSs experience relative deflection in the direction towards the LMC. HVS located between $-20$ kpc $ < z < 20$ kpc exhibit a similar pattern of deflection to each other. Stars located at $z> + 20$ kpc show opposite trends to their below-disc counterparts. This behavior is illustrated more clearly in the $y-z$ plane, which reveals a symmetric cone-like structure where stars are deflected towards $+\hat{\theta}$ if they are either a) on trajectories which would take them `underneath' the LMC, or b) on small-$\theta$ trajectories heading away from the LMC, and towards $-\hat{\theta}$ otherwise. Overall, HVS are preferentially deflected towards the LMC in both $\beta_\phi$ and $\Delta \beta_\theta$. Additionally, the overall magnitude of the deflections is greater than that of the bar. \par 

Fig. \ref{fig:beta_nif} shows that the MW+NIF potential overall has the opposite influence on HVSs than the MW+LMC potential. We see that HVSs which had positive values of $\beta_\phi$ and $\Delta \beta_\theta$ in the MW+LMC potential have the opposite signs in the MW+NIF potential. This is due to the effect of the GC barycenter accelerating towards the LMC's location due to the mass of the LMC itself. This non-inertial frame effectively gives the HVSs an apparent extra component of velocity in the opposite direction of the LMC, translating to a reversal of trends between the two potentials. \par 

Lastly, Fig. \ref{fig:beta_all} plots the deflection of the HVSs sample through the MW+All potential. Here, we look at the combined influence of the bar, LMC, and reflex motion. We find that the influence of the bar is primarily limited to the region close to the GC. We see that close to the GC in the first row of the $\beta_\phi$ plots, the spiral distorts the central region of the $-20$ kpc $< z < 0$ kpc and $0$ kpc $ < z < +20$ kpc panels. The bar effect is also greater for HVSs above the disc since the LMC influence lessens the far above the disc. The bar influence is competing with both the LMC and the NIF, which makes it difficult to disentangle. The LMC's direct influence is strongest for HVSs ejected below the disc, and that are currently located at $z < -20$ kpc. In the $x-y$ projection for $\beta_\phi$, the LMC has a defined sphere of influence within which HVSs are deflected towards the LMC itself. Outside of this sphere, the reflex motion effects dominate. HVSs located between $-20$ kpc $ < z < 0$ kpc still shows a mix of LMC and NIF influence, but the defining line between them is not as sharp. The same is true of the HVS between $0$ kpc $ < z < +20$ kpc. The $z > +20$ kpc in the $x-y$ projection almost completely recovers the MW+NIF behavior. In the $y-z$ projection, the $\beta_\phi$ behavior, we once again see that the LMC influence in $\beta_\phi$ dominates only for stars ejected towards its general location, i.e. towards $-y$ and $-z$, with bar effects apparent near the GC and the NIF effects dominating elsewhere. The story for $\Delta \beta_\theta$ is similar, where we see that HVSs located close to the LMC are deflected towards its position, but the NIF influence dominates elsewhere.

\subsection{HVS Observability}\label{sec:results-obsv}

Our analysis has shown that both $\beta_\phi$ and $\Delta \beta_\theta$ are typically quite small regardless of the potential model---only up to $\sim$one degree in magnitude and often several orders of magnitude smaller. Measuring these angles from Earth in practice would require extremely precise measurements of the HVS kinematics. To explore this further, we take only stars detectable in the Gaia DR4 radial velocity catalog and estimate their astrometric and radial velocity observational uncertainties based on the Gaia DR4 predicted performance (see Section \ref{sec:methods-select-sample} for details). In this detectable subsample, we only take stars with estimated relative parallax uncertainties less than 20 per cent since these would stand out as promising HVS candidates in the data release---geometric distance estimation becomes non-trivial for larger relative uncertainties \citep[see][]{BailerJones2015}. After applying these magnitude and precision cuts to the HVSs in the MW+All sample, we are left with 234 observable HVSs out of the 23,409 unbound stars in our mock catalog. Of these, we take the 40 stars with effective temperatures cooler than 14500 K, for whom heliocentric radial velocity uncertainties are straightforward to model with \texttt{pygaia}. We then sample 1000 times over the estimated errors in parallax, proper motion and radial velocity and repeatedly calculate $\beta_\phi$ by transforming the sampled kinematics into Galactocentric coordinates. We focus only on the azimuthal deflection $\beta_{\phi}$ to determine whether we can observe the impact of the non-axisymmetric potential components. The uncertainty in the MW-only contribution to $\Delta\beta_\theta$ makes it even more difficult to measure and so we limit our observability analysis to $\beta_\phi$ only. If the spread in the sampled $\beta_\phi$ is small compared to the real $\beta_\phi$, then we could expect Gaia DR4 to provide a measurement for $\beta_\phi$ for the star with reasonable confidence.

\begin{figure}
    \centering    
    \includegraphics[width=\columnwidth]{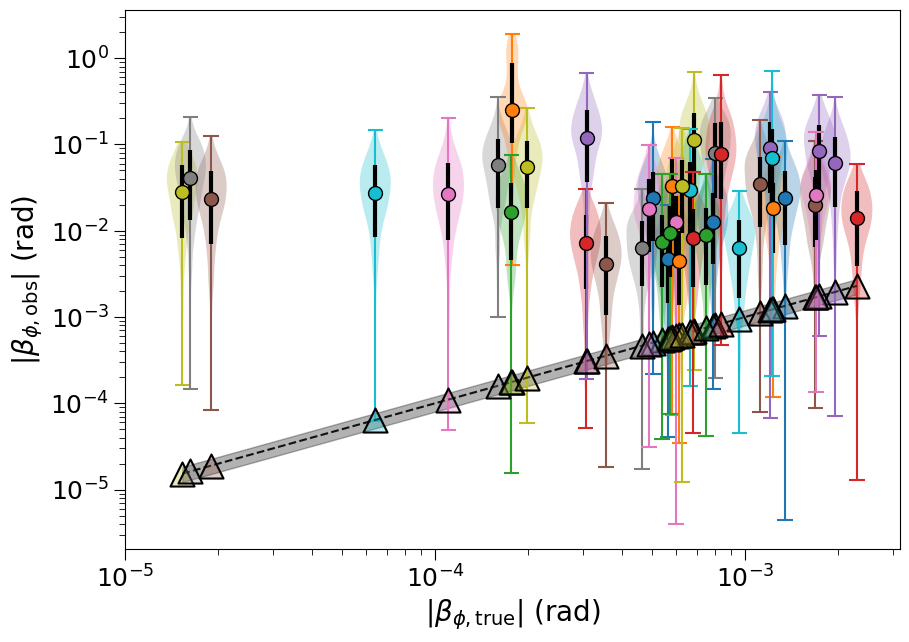}
    \caption{Violins show the range of measurements of $|\beta_\phi|$ when taking into account observational uncertainties in a subsample our MW+All sample selected to be detectable in Gaia DR4 (see text for details). The coloring is arbitrary and serves to differentiate the violins. The extrema of the violins show the minimum and maximum observed $|\beta_\phi|$ over 1000 iterations, while the  circles show the median and the black bars span the 16th to 84th percentiles. The triangles and dashed line show the 1:1 line where the measured $|\beta_\phi|$ matches the true $|\beta_\phi|$ and the shaded region shows $\pm$20 per cent above and below this line.}
    \label{fig:BetaPhiViolins}
\end{figure}

Fig. \ref{fig:BetaPhiViolins} is a violin plot of this process on a subset of 40 of the observable HVSs, where the triangles show the true $|\beta_\phi|$ for each mock HVS and the violins show the distribution of measured $|\beta_\phi|$ upon sampling the errors. The range in measured $|\beta_{\phi}|$ for the observable subsample of HVSs spans several orders of magnitude ranging from as low as $\sim10^{-6}$ rad and in some cases as large as $\pi$ rad.
The median measured $|\beta_\phi|$ is typically several orders of magnitude larger than the true $|\beta_\phi|$ and only very rarely does a measurement of $|\beta_\phi|$ approach the true value to within a factor of a few. Sampling over the observational uncertainties, the probability of measuring a $|\beta_\phi|$ within 20\% of the true value is at most a few per cent. 

\par 
To understand further how different sources of measurement error impact the ability to reliably measure $\beta_\phi$, we assume the declination position, parallaxes, proper motions, and radial velocities of these HVSs are known to perfect precision and only the right ascension positions are uncertain. We repeat the sampling process for each of the observable HVSs and determine the uncertainty in measured $|\beta_\phi|$, taken as half of the 16th to 84th percentile range. We then set the right ascension uncertainty to zero and repeat this process five more times, allowing the declination, then parallax, then right ascension proper motion, then declination proper motion, then radial velocity all to vary one at a time. 

\begin{figure}
    \centering
    \includegraphics[width=0.9\linewidth]{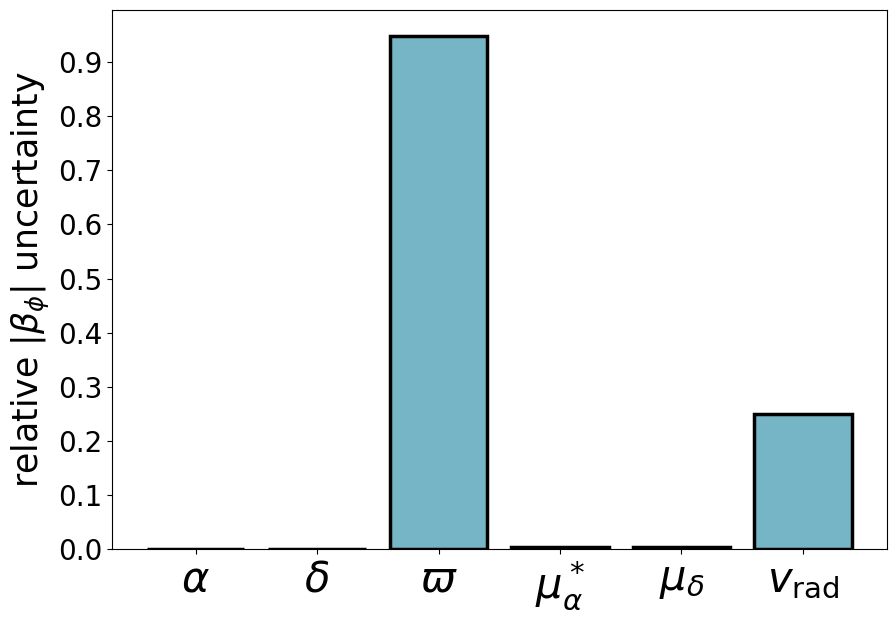}
    \caption{Mean uncertainty in measured $|\beta_\phi|$ when only one kinematic variable has uncertainties and the rest are known to perfect accuracy, expressed as a fraction of the mean $|\beta_\phi|$ uncertainty when all variables have uncertainties. Relative uncertainties are shown when the right ascension ($\alpha$), declination ($\delta$), parallax ($\varpi$), proper motion in each direction ($\mu_\alpha^*, \;\mu_\delta$) and heliocentric radial velocity ($v_{\rm rad}$) are the isolated variables.}
    \label{fig:errors_barchart}
\end{figure}

The results of this test can be seen in Fig. \ref{fig:errors_barchart}, where we show these recalculated uncertainties in $|\beta_\phi|$, scaled relative to the uncertainty when all variables are allowed to vary. When parallax is the only uncertain variable, the uncertainty in observed $|\beta_\phi|$ is nearly as large as when all observables are varied. If the line-of-sight velocity is the only component allowed to vary, the observational uncertainty of $|\beta_\phi|$ is reduced by $\sim$75\%. If either proper motion component is the only source of uncertainty, the observed $|\beta_\phi|$ uncertainty is reduced by a factor of 100 and there is a $>$50\% chance of the measured $|\beta_\phi|$ being within 20 per cent of the true $|\beta_\phi|$ for about half of the observed stars. Finally, if the already extremely precise sky position of the star is the only uncertainty, $|\beta_\phi|$ uncertainties are reduced by a factor of $10^8$ and the true $|\beta_\phi|$ is always recovered. In summary, this simple test shows that parallax uncertainty is mainly responsible for our inability to measure $|\beta_\phi|$ today and in the near future. While spectroscopic follow-up observations could in principle offer significantly improved radial velocity errors, prospects for improving distance/parallax estimates are not particularly promising for the proposed next generation of Galactic astrometric surveys. In Gaia's ten-year end-of-mission data release 5, parallax uncertainties will improve by a factor of $\sqrt{2}$ and the proper motions by a factor of 2$\sqrt{2}$. If the proposed Gaia successor mission GaiaNIR \citep{Hobbs2016} also completes a ten-year survey with twenty years between the missions, eventually proper motions will improve by a factor of 20 relative to Gaia DR4 but parallaxes by only a factor of 2. Given these astrometric improvements and assuming a radial velocity uncertainty of $0.1 \; \mathrm{km \ s^{-1}}$, parallax uncertainties would \textit{still} need to improve by an additional factor of 150 for the probability of measuring $|\beta_\phi|$ within 20 per cent of the true value to reach $>$50\% for at least half the DR4-detectable HVSs. If follow-up observations were to obtain independent non-trigonometric distance measurements for these HVSs, the uncertainties would need to be better than 0.03\% to achieve this same performance. To compound the issue, even if measurement precisions like this were achieved, accurate measurements of $\beta_\phi$ would additionally require very precise determinations of the distance between the Sun and the GC \citep[c.f.][]{Leung23} and the velocity of the Sun in the Galactocentric rest frame.

\section{Discussion} \label{sec:discussion}
We find that all four of our potentials of interest (MW+Bar, MW+LMC, MW+NIF, and MW+All) cause HVSs to deviate from the trajectories they would take in an axisymmetric MW potential and would severely hamstring efforts to use them to constrain the Galactic potential. While we have stuck to describing general trends, the exact and quantitative nature of some of these observed trends depend on the assumptions we have made about aspects of the Galactic potential and about the HVS ejection mechanism. The initial mass function in the GC and the distribution of GC binary separations and mass ratios all impact the HVS mass and velocity distribution to varying degrees \citep[see][]{Evans2022b}. Our work additionally looks only at HVSs ejected via the \cite{Hills1988} mechanism. Alternative HVS ejection scenarios in the GC include interactions between single stars and a binary consisting of Sgr A* and an as-of-yet-undetected intermediate-mass or massive companion black hole \citep{Yu2003, Sesana2006, Levin2006, Rasskazov2019, Evans2023}, or interactions with a sizable population of stellar-mass black holes in the GC \citep{OLeary2008}, or the tidal disruption of an infalling globular cluster \citep{Capuzzo2015, Fragione2016a}. Each of these mechanisms would have its own characteristic HVS mass and velocity distribution, which would impact the precise degree to which the bar and LMC impact their trajectories. 

Throughout this work we assumed that all the baryonic components of the Galactic potential were fixed when we were assessing the likelihoods of different DM halos, and we assumed all static components of the potential were fixed when we were exploring $\beta_\phi$ and $\beta_\theta$. There exists significant ongoing uncertainty over the parameters which describe the Galactic potential \citep[see][]{Bland-Hawthorn2016}, particularly for the DM halo but including the baryonic components as well. For example, our MW-only potential has a rather heavy bulge compared to other widely-used MW potentials such as \texttt{MWPotential2014} \citep{Bovy2015} or the best-fit potential of \citet{McMillan2017} and as such has a larger escape velocity in the inner regions of the Galaxy. The \citet{McMillan2017} DM halo, however, is more massive and its escape velocity is larger than MW-only's beyond 2 kpc. While the overall trends shown in this work are not particularly sensitive to these choices in the potential, heavier axisymmetric components will act as a moderating influence, decreasing the degree to which the non-axisymmetric components influence HVS kinematics.

For that matter, we have also assumed a fixed LMC mass of $M_{\rm LMC} = 1.5\times 10^{11} \; \mathrm{M_\odot}$ \citep{Erkal2019LMCmass}. However, there is still some uncertainty surrounding the exact value, and a range of values between $1.1\times10^{11} \; \mathrm{M_\odot} - 2.5\times10^{11} \mathrm{M_\odot}$ are reported in the literature \citep[see][]{Penarrubia2016, Erkal2020LMCmass, Watkins2024}. While a significant change to the LMC mass would not impact the qualitative trends in the MW+LMC and MW+NIF potentials explored here, it would change their amplitudes and the degree to which they interfere with the effects of the bar in the MW+All potential. The same can be said of our fixed assumptions for the bar pattern speed, position angle, strength and length of the bar in Sec. \ref{sec:methods:potential}, though they are rather less contentious and vary only at the $\sim$10 per cent level in contemporary studies \citep[see additionally][]{Sanders2019, Clarke2022, Li2022bar, Xia2024}. Regardless, variations in the bar potential parameters would affect the strength of $\beta$ deflection and the degree of winding in Figs. \ref{fig:beta_bar} and \ref{fig:beta_all}.
\par 

We show that the potential constraining methods of \citetalias{Contigiani2019} and \citetalias{Gallo2022} fail when applied to any of the non-axisymmetric potentials. These methods were designed and tested on axisymmetric MW potentials similar to our MW-only. This suggests that potential constraining methods are not immune to the small, but important, influence these non-axisymmetric and time-dependent components have on HVS trajectories. One avenue for future work is to determine if the above methods could be adapted to  account for non-equilibrium components of the potential and simultaneously constrain, for example, not only the Galactic DM halo but also the mass and size of the LMC and the length, pattern speed and strength of the bar. This, however, would add even more dimensionality to an already multi-dimensional, degenerate, and computationally expensive venture.

When looking at the deflection of HVS in response to the LMC, 
our results align with \citet{Boubert2020}. They find a difference in the angle between the Galactocentric position and velocity vectors of HVSs propagated through a Galactic potential containing the LMC of up to $3^\circ$, or $\sim 5\times10^{-2}$. Additionally, we further find similar spatial trends, where HVS are deflected towards the LMC, with the magnitude of the deflection depending on the HVS-LMC distance. Our work builds on their conclusion that the LMC and associated reflex motion are import considerations when constraining HVS trajectories by further demonstrating that the Galactic bar must also be accounted for. We find this particularly important for HVSs which are still close kpc to the GC, which is most clearly shown in the middle two columns of Fig. \ref{fig:beta_all}. 
\par 

In this work we have limited our analysis to the Galactic bar, LMC and associated effects. However, these are not the only non-axisymmetric components of the Galaxy. Other potential sources of non-axisymmetry and time dependence in the Galaxy which could influence HVS trajectories include the contribution of spiral arms to the Galactic potential \citep[see][]{Cox2002, Han2014}, non-axisymmetry due to the boxy/peanut mass distribution within the bulge \citep[see][]{Gonzalez2016}, or perturbations from giant molecular clouds near the GC \citep[see][]{Morris1996, Walls2016, Rogers2022}. Each of these features contribute to the Galactic potential and are structures which could be capable of deflecting HVSs. This work has shown that even seemingly small components of the Galactic gravitational potential are capable of affecting HVS trajectories in a meaningful way, and so further investigation into other sources of non-axisymmetry and time dependence in the Galactic potential is warranted. \par 

\section{Conclusions} \label{sec:conclusions}

Hypervelocity stars (HVSs) are an exciting class of star whose high velocity makes them unbound to the Galaxy. They are a unique tool which could probe the structure of the Milky Way's gravitational potential. The trajectory of HVSs as they escape the Milky Way contains information about the potential through which they are traveling. \par 

In this work we show that the Galactic bar and LMC are important components of the Milky Way's gravitational potential and should be considered when trying to use HVSs to constrain Galactic potential. We conclude that:
\begin{enumerate}
    \item The inclusion of the Galactic bar, LMC and associated reflex motion causes HVSs propagated backwards in time through axisymmetric potentials to potentially miss the GC by up to several hundred parsecs (Figs. \ref{fig:BarLMCHist} and \ref{fig:BarLMCAngMom}).
    \item Methods proposed by \citet{Contigiani2019} and \citet{Gallo2022} using hypervelocity stars and simple models of the Galactic potential to constrain the Galactic dark matter halo are unsuccessful when the Galactic bar and the LMC and its associated reflex motion are considered. While these methods work for our fiducial MW-only potential, the presence of even one of these effects is enough for potential reconstruction attempts to fail (Figs. \ref{fig:likelihood} and \ref{fig:Gallo2D}). 
    \item Deflection of the HVSs is quantified using $\beta$, the difference between the HVSs position and velocity vector. 
    We find that different components dominate in different regimes, with the bar exerting the strongest influence on HVS located close to the Galactic mid-plane and near to the bar itself. The LMC exerts the greatest influence close to its location, mostly impacting HVS located at $z < -20$ kpc and within $\sim 50$ kpc of the LMC. Everywhere else we find that the non-inertial frame force generated by the reflex motion of the GC in response to the passing LMC dominates (Figs. \ref{fig:beta_bar} - \ref{fig:beta_all}).
    \item Mock observations using Gaia DR4 predicted performance show that $\beta_\phi$, the component of $\beta$ in the azimuthal direction is too small to reliably measure in the upcoming data release. Uncertainty in parallax contributes the most to the measurement uncertainty, and so significantly improved distance estimates in the future are required to measure $\beta_\phi$ in practice (Figs. \ref{fig:BetaPhiViolins} and \ref{fig:errors_barchart}).
\end{enumerate}

Our results suggest that these non-axisymmetric effects must be considered when using HVSs to constrain the Galactic potential, and that potential fitting models must account for the appreciable angular momentum imparted by the bar, LMC, and reflex motion. We have shown that when it comes to the Galactic potential, one must indeed sweat the small stuff -- even seemingly inconsequential components of the Galactic potential are capable of drastically impacting potential fitting attempts when not properly accounted for. 

\section*{Acknowledgments}
The work of IA was supported in part by an NSERC USRA. FAE acknowledges support from the University
of Toronto Arts \& Science Postdoctoral Fellowship and Dunlap Postdoctoral Fellowship programs. JB acknowledges financial support from NSERC (funding reference number RGPIN-2020-04712).

\section*{Data Availability}
The simulations underpinning this work can be provided upon reasonable request to the corresponding author. The \texttt{python} package \texttt{speedystar} \citep{Contigiani2019, Evans2022b} was used to generate the simulations and is publicly available at \url{https://github.com/speedystar}. Potential modeling was done using \texttt{galpy} \citep{Bovy2015}, available at \url{ http://github.com/jobovy/galpy}. Python packages used for data analysis and visualization include \texttt{astropy}  \citep{astropy:2018}, \texttt{matplotlib} \citep{Matplotlib2023}, and  \texttt{numpy} \citep{Harris2020}.

\bibliographystyle{mnras}
\bibliography{HRS}

\end{document}